# Anomalous Thermal Conductivity Reversal Driven by a Fluctuating Charge Density Wave

Han Huang[1]†, Jinghang Dai[1]†, Joyce Christiansen-Salameh[1], Jiyoung Kim[1], Samuel Kielar[1], Desheng Ma[2], Noah Schinitzer[3,4], Danrui Ni[5], Gustavo Alvarez[1], Chen Li[1], Carla Slebodnick[6], Mario Medina[1], Bilal Azhar[1], Ahmet Alatas[7], Robert Cava[6], David Muller[2,4], Zhiting Tian[1,*]

[1]Sibley School of Mechanical and Aerospace Engineering, Cornell University, Ithaca, New York 14853, USA

[2]School of Applied and Engineering Physics, Cornell University, Ithaca, NY 14853, USA

[3]Department of Materials Science and Engineering, Cornell University, Ithaca, NY 14853, USA

[4]Kavli Institute at Cornell for Nanoscale Science, Cornell University, Ithaca, NY 14853, USA

[5]Department of Chemistry, Princeton University, Princeton, New Jersey 08544, USA

[6]Department of Chemistry, Virginia Tech, Blacksburg, Virginia 24061, USA

[7]Advanced Photon Source, Argonne National Laboratory, Argonne, Illinois 60439, USA

†These two authors contribute equally

*Corresponding author. Email: zhiting@cornell.edu

**Abstract:** For more than a century, the lattice thermal conductivity of crystalline solids has been known to decrease with increasing temperature, as the phonons that carry heat scatter more strongly via Umklapp processes. Here we report a pronounced violation of this rule in the layered charge-density-wave (CDW) metal 2H-$TaSe_2$. Its in-plane thermal conductivity is strongly nonmonotonic: it peaks sharply at the 122 K ordering transition, collapses to a deep minimum near 210 K, and then anomalously reverses, rising with temperature to at least 350 K. Such a high-temperature reversal, in which a crystal conducts heat more efficiently as it is heated, is rare in crystalline solids. Its thermal conductivity minimum lies deep in the disordered phase well above the ordering transition, unlike the weak, transition-centered anomalies of other charge-density-wave systems, pointing to heat conduction governed by fluctuating rather than established order. Combining inelastic X-ray scattering, selected-area electron diffraction, and synchrotron X-ray diffraction, we attribute the reversal to phonons scattering off persistent, fluctuating charge order that survives more than 150 K above the long-range transition, to at least 300 K. A phenomenological phonon–CDW scattering model, tied directly to the measured diffraction linewidth and intensity, quantitatively reproduces the entire nonmonotonic curve. We also reveal that the charge order melts not abruptly but through a fluctuation- and defect-driven, weak first-order process, leaving short-range correlations far above the transition. This work establishes fluctuating charge order as a sensitive control knob for thermal transport in quantum materials and thermal conductivity as a powerful probe of the hidden charge order.

## INTRODUCTION

The thermal conductivity of a crystalline solid is among its most fundamental properties, and its temperature dependence is among the most predictable. Above the low-temperature peak, the phonons that carry heat scatter more strongly as the lattice warms, so the conductivity falls monotonically with temperature. Identified experimentally by Eucken more than a century ago and explained by Peierls through Umklapp phonon scattering, this decrease is observed in nearly all crystalline insulators and semiconductors, with rare exceptions in magnetic and strongly anharmonic crystals[1–4], and ranks among the most robust rules in solid-state physics. Quantum materials are another matter. With their electronic, lattice, and topological degrees of freedom strongly intertwined, they repeatedly defy the rules that govern ordinary crystals, yet whether this propensity for rule-breaking reaches heat conduction has drawn surprisingly little attention. The layered transition-metal dichalcogenides (TMDCs) are an ideal place to look: quasi-two-dimensional quantum materials in which charge, lattice, and electronic orders compete and coexist.

Among these TMDCs, 2H-$TaSe_2$ is a canonical layered charge-density-wave (CDW) material that, on cooling, undergoes an incommensurate CDW transition at 122 K followed by a commensurate transition at 90 K [5,6]. Despite decades of study using spectroscopic [7–13], transport [14–16], scattering [6,7,17–21], and real-space techniques [22,23], key aspects of the CDW transition nature remain controversial. Several observations challenge traditional mechanisms involving Fermi surface nesting or saddle points: the metallic optical conductivity below the transition temperature [24], the large CDW gap (~0.1 eV) compared to $T_{CDW} = 122\,K$ [25,26], and the short coherence length ($\xi^0 \sim 3 - 10$ Å) indicating strong fluctuations [27]. Real-space transmission electron microscopy (TEM) studies of 2H-$TaSe_2$ crystals have also shown stripe-phase nucleation and dislocation-mediated evolution [22], which underscore the essential microscopic role of defects. The CDW order in 2H-$TaSe_2$ is thus shaped by an intertwined interplay of fluctuations, defects, and lattice dynamics that largely eludes electronic and optical probes, which do not couple directly to the charge-neutral lattice distortion.

Heat-carrying phonons couple directly to these lattice distortions. We measured the in-plane thermal conductivity of high-quality 2H-$TaSe_2$ single crystals (Fig. 1(d)) from 78 to 350 K using transient thermal grating (TTG) spectroscopy, a non-contact optical technique. The result is striking: instead of the usual monotonic decrease, the thermal conductivity is strongly nonmonotonic, peaking sharply at the 122 K transition, falling to a deep minimum near 210 K, and then anomalously reversing to rise with temperature to the highest temperatures measured. A thermal conductivity that increases on heating runs counter to the behavior of common crystalline solids. Inelastic X-ray scattering, selected-area electron diffraction (SAED), and synchrotron X-ray diffraction (XRD) trace this anomaly to short-range charge density wave order that survives far above the long-range transition, to at least 300 K. This residual order scatters phonons and melts not abruptly but through a fluctuation- and defect-driven, weak first-order process. A phenomenological model that links phonon–CDW scattering to the measured diffraction linewidth and intensity quantitatively reproduces the full nonmonotonic curve from our TTG measurements.

Such a signature trend emerging in a bulk transport property is significant in itself. Our results further identify a fluctuating, charge-neutral order as a previously unexplored control knob for heat flow in low-dimensional quantum materials, opening a new route toward thermal switching. Because phonons couple to virtually every excitation in a crystal, thermal conductivity is acutely sensitive to collective modes and hidden dynamics that elude conventional spectroscopies. The argument runs in reverse as well: the very fact that a charge-neutral, fluctuating order leaves such a distinct imprint on heat conduction establishes thermal transport as a sensitive probe of the complex, fluctuation-driven phase transitions that are otherwise hard to reach. Beyond being a property to be explained, the thermal conductivity thus becomes a tool for probing exotic quantum phases. Intriguingly, the fluctuating local CDW we observe mirrors the high-temperature fluctuating charge order of cuprate high-Tc superconductors [28,29], hinting at universal features in the physics of low-dimensional quantum materials.

## METHODS

2H-$TaSe_2$ has a hexagonal (*P*$6_3$/*mmc*) structure consisting of two layers of Se–Ta–Se sandwiches stacked along the c-axis, shown in **Fig. 1(a)**. Single crystals were synthesized via a two-step solid-state reaction followed by iodine-assisted chemical vapor transport in a temperature gradient of 1050–1000 °C for one week [30]. The high crystal quality was verified by structural characterization: single-crystal X-ray diffraction confirmed the pure 2H phase [31–34], and atomic-resolution high-angle annular dark-field scanning transmission electron microscopy (HAADF-STEM) imaging resolved the ordered atomic lattice of a clean single crystal (Fig. 1(d)). A microscope image of the measured sample is shown in Fig. 1(b). IXS was performed at the Advanced Photon Source,

Argonne National Laboratory [35,36] to probe lattice dynamics along the [100] and [110] directions at 300 K, 115 K, and 80 K. The IXS data were fitted with the damped harmonic oscillation model to extract the phonon frequency and linewidth [31–34]. XRD was performed at the Cornell High Energy Synchrotron Source (CHESS) from 80 K to 130 K with both cooling and warming cycles to detect thermal hysteresis. First-principles calculations using Temperature Dependent Effective Potentials (TDEP) [37,38] with Vienna Ab Initio Simulation Package (VASP) [39,40] were performed at room temperature to obtain finite temperature lattice dynamics (details in Supplementary Information (**SI**)).

In-plane thermal transport was measured using the TTG technique, a non-contact optical pump–probe method illustrated in **Fig. 1(c)** [41–46] (details in **SI**). Exfoliated crystals were mounted on a coin-like copper holder with silver paste to ensure clean laser reflection and good thermal contact at cryogenic temperatures. TTG data were fitted to extract the in-plane thermal diffusivity from 77 K to room temperature. (**Fig. S1**). Thermal conductivity from 78 K to 350 K was then obtained by combining the diffusivity with the specific heat estimated using the Debye model [47] (see **SI**), as shown in **Fig. S2,** and the density determined from XRD.

SAED measurements were performed on a mechanically exfoliated 2H-TaSe2 flake transferred via polypropylene carbonate (PPC) stamp to a silicon nitride sample support. Diffraction patterns were collected on a Thermo Fisher Scientific Spectra 300 X-CFEG operated at 120 kV with a liquid nitrogen-cooled Gatan 636 specimen holder, and sample temperature was controlled with a Gatan 1905 heating controller. Atomic-resolution HAADF-STEM images (Fig. 1(d)) were acquired on the same microscope operated in scanning mode.

## RESULTS AND DISCUSSION

Guided by the premise that phonons couple broadly to lattice distortions, electronic ordering, and topological defects, we first examine the temperature dependence of the thermal conductivity across the CDW transition. As shown in **Fig. 2**, the normal phase (> 122 K) thermal conductivity exhibits a distinctive V-shaped temperature dependence. A sharp peak appears at $T_{CDW} = 122\ K$, coinciding with the incommensurate to normal CDW transition. This peak originates from a $\lambda$-shaped specific heat peak at critical temperature driven by critical-like fluctuations of the CDW order, a hallmark of continuous phase transitions (SI). Upon warming, the thermal conductivity decreases until reaching a minimum around 210K. Most remarkably, above 210K, we observe an anomalous increase in thermal conductivity that persists to high temperatures, highly unusual for crystalline materials, where thermal conductivity typically decreases with increasing temperature due to enhanced phonon-phonon scattering. Meanwhile, the electronic contribution estimated from the Wiedemann-Franz law is small (around 5 $W/m \cdot K$) [48] and does not account for the observed anomalous trend. This reversal is the central finding of this work: a crystalline solid that conducts heat more efficiently the hotter it becomes, in defiance of conventional phonon transport.

To assess whether changes in lattice dynamics alone can account for this behavior, we turn to IXS measurements. At 115 K, the IXS data reveal distinct low-energy modes near the CDW wave vector $\boldsymbol{q}_{CDW} \approx (0.33, 0, 0)$, at $q = 0.3$ and $q = 0.4$ along the $\Gamma - M$ direction (**Fig. 3**) compared to 300 K. No discernible phonon peak was observed at $q = 0.35$, likely because the signal is masked by strong elastic scattering when $q$ is very close to $\boldsymbol{q}_{CDW}$. Comparison with prior IXS measurements from the same beamline [18] suggests that these modes arise from phonon softening. However, since these low-energy points disappear at 80 K, we do not rule out that they represent phase excitations specific to the incommensurate phase. The associated band bending near $q_{CDW}$ results in a steep dispersion slope and higher phonon group velocity ($v_{\boldsymbol{q}} = \partial_{\boldsymbol{q}}\omega_{\boldsymbol{q}}$), which may contribute to the conductivity enhancement near the transition (from 122 K to ~210 K). However, the enhanced phonon group velocity is confined to a narrow momentum region ($\delta|\boldsymbol{q}|$~0.1) [18], rendering it insufficient to account for the magnitude and breadth of the thermal conductivity change from 122 K to ~210 K. Also, this mechanism cannot account for the anomalous recovery of thermal conductivity observed at high temperatures above ~210K. Moreover, the recovery of thermal conductivity above 210 K does not coincide with any known structural or electronic transition, indicating that conventional phonon-phonon scattering alone cannot explain the observed behavior, because upon warming, phonon-phonon scattering becomes stronger and suppresses thermal conductivity. Instead, these observations point to a competition between distinct phonon scattering channels, motivating consideration of unconventional scattering mechanisms associated with fluctuating order and defect-mediated lattice distortions at elevated temperatures. Conventional anharmonic phonon–phonon scattering, which only intensifies on warming, is thereby excluded as the origin of the high-temperature reversal, pointing instead to a fluctuation-driven mechanism.

To test this hypothesis, we examine how the underlying structural correlations evolve in the high-temperature regime. Our SAED measurements reveal distinctive residual CDW superlattice peaks persist even at 300K (**Fig. 4**), far above the long-range-order transition temperature of 122 K. These surviving satellite peaks demonstrates robust short-range (local) CDW order deep into the nominally disordered phase. We also observe a linear increase in SAED peak widths with increasing temperature, as shown in **Fig. 4(e)**, suggesting a decreasing coherence length with $\frac{1}{|T-T_{CDW}|^{\eta}}$ scaling behavior, where $\eta \approx 1$. Notably, this scaling is valid not just a few Kelvins near the critical temperature but persists more than 150 K above $T_{CDW}$, underscoring the unusually long-lived nature of CDW correlations in this system. Importantly, the loss of CDW coherence manifests primarily through the suppression and broadening of discrete satellite peaks, rather than emerging into a ring-like diffuse scattering. The absence of pronounced azimuthal broadening suggests that the high-temperature CDW state does not correspond to a hexatic phase, as would be expected in a canonical KTHNY melting scenario, where orientational order is lost prior to translational order. Instead, in 2H-$TaSe_2$, the CDW wavevector remains locked to in-plane crystallographic directions even as translational coherence is progressively reduced. Previous XRD study on 2H-$TaSe_2$ also reported similar broad, nearly isotropic diffuse peaks persisting up to 200K [49]. We attribute the absence of a diffuse ring to dimensionality and orientational pinning: in our bulk 2H-$TaSe_2$, interlayer coupling and lattice locking fix the CDW direction to in-plane crystallographic axes, suppressing the azimuthal averaging prominent in truly two-dimensional monolayers (e.g., 1T-$TaS_2$ [50]), thereby yielding fading lobes rather than a ring.

While SAED exposes the fluctuation-dominated character of the high-temperature phase, synchrotron XRD supplies a complementary thermodynamic view of the transition. Consistent with prior studies, our XRD confirms the normal-to-incommensurate and incommensurate-to-commensurate CDW transitions at 122 K and 90 K, respectively (Fig. 5). Decisively, the CDW wavevector exhibits a clear thermal hysteresis between cooling and warming cycles near the incommensurate transition (Fig. 5(c)). This hysteresis is direct thermodynamic evidence of latent energy, a defining signature of a first-order transition.

Taken together, the unconventional phonon scattering above $T_{CDW}$, the divergent specific heat at the critical point, the persistence of short-range CDW order far above $T_{CDW}$ revealed by SAED, and the clear hysteresis of the CDW wavevector point to a weak first-order melting scenario for the CDW transition in 2H-$TaSe_2$, which has not been previously recognized in CDW systems. In this regime, the transition involves a subtle but finite discontinuity, manifested through latent energy and phase coexistence, while remaining strongly influenced by fluctuations and defect proliferation and giving rise to an extended temperature range of short-range order. This behavior is fundamentally distinct from both conventional first-order transitions, which exhibit abrupt loss of order with minimal precursor fluctuations, and continuous KTHNY-type melting [51,52], in which translational order is lost through the continuous unbinding of topological defects, with neither latent heat nor thermodynamic hysteresis. Our observations instead place 2H-$TaSe_2$ in the intermediate regime anticipated by theory, where a genuine, if subtle, discontinuity coexists with strong fluctuations; this picture aligns with contemporary frameworks for transitions in 2D and quasi-2D systems that admit multiple melting routes depending on the microscopic parameters and interaction potentials[53–57]. Notably, related 2D CDW materials such as 1T-$TaS_2$ also exhibit persistent local CDW order well above their transition temperatures [50,58,59], suggesting that dislocation- and fluctuation-driven melting may represent a unifying description for CDW transitions in low-dimensional layered materials, but the melting in 1T-$TaS_2$ is of KTHNY type involving a hexatic process. Our results also argue against the Ising-type transition model proposed by Gor'kov [60], in which Ta atoms hop between two potential minima in analogy with spin flips due to strong electron-phonon coupling. In this Ising scenario, the transition is strictly second order at $T_{CDW}$. While strong electron-phonon coupling remains central to the ordered CDW phase, evidenced by phonon softening, the transition itself proceeds through a melting-like evolution characterized by continuous spectral broadening and q-vector hysteresis.

The persistent short-range CDW order above $T_{CDW}$ provides a natural explanation for the anomalous thermal conductivity. As temperature increases through $T_{CDW}$, the long-range CDW order gradually melts into short-range correlations, as shown by the reduced coherence length and persistent satellite peaks in SAED. In this intermediate regime, growing spatial fluctuations enhance phonon-CDW scattering, leading to suppressed thermal conductivity. Meanwhile, as the temperature increases, the local CDW domains become increasingly short-ranged and dilute, resulting in fewer phonon-CDW scattering events. This competition between these two effects explains the unusual V-shaped temperature dependence observed in our thermal conductivity measurements, with a minimum around 210K separating the two competing regimes. Once the CDW amplitude collapses at still higher temperatures, this scattering channel switches off, the phonon mean free path recovers, and the thermal conductivity rises, which accounts naturally for the anomalous high-temperature reversal.

This picture can be formalized by modeling phonon scattering from spatial fluctuations of the CDW order parameter (full details in **SI**), expressed as $\psi(r)a = A(r)\exp\left[iq\left(z + u(r)\right)\right]$, where $A(r)$ is the CDW amplitude and $u(r)$ represents wavefront distortions. The spatial gradient of the order parameter ($\nabla\psi(r)$) quantifies disorder caused by elastic deformations and dislocations. Within the Born approximation, the scattering rate is proportional to the variance of these gradients, given by

$$\tau_{CDW}^{-1}(T) \propto \langle|\nabla A|^2\rangle + q^2\langle A^2(\nabla u)^2\rangle. \qquad (1)$$

Experimentally, the CDW diffraction peak width $w(T)$ reflects $\langle(\nabla u)^2\rangle$, while the integrated peak intensity tracks $A^2(T)$. Thus, we can show that the CDW contribution to the scattering rate can be expressed phenomenologically as

$$\tau_{CDW}^{-1}(T) = a w(T)^2 A^2(T), \qquad (2)$$

with $a$ being a material-dependent constant. The total thermal conductivity is then obtained from kinetic theory as $\kappa(T) = \frac{1}{3}C_v(T)v_s^2\tau_{tot}(T)$, where $v_s$ is the sound velocity and $C_{v(T)}$ is the phonon specific heat per unit volume. The scattering rate is given by a Matthiessen sum

$$\tau_{tot}^{-1}(T) = A_U T\exp\left(-\frac{\Theta_U}{T}\right) + A_{imp} + \tau_{CDW}^{-1}(T) \qquad (3)$$

where $A_U$ and $\Theta_U$ parametrize Umklapp scattering, and $A_{imp}$ accounts for impurity scattering. The specific heat is described by the Debye model with $\Theta_D = 200\,K$. The peak width is fitted as $w(T) = w^0 + m(T - 100K)$, where the fitted $w_0 \approx 0.0865\,\text{Å}^{-1}$ and $m \approx 0.00168\,\text{Å}^{-1}K^{-1}$. The amplitude softening is modelled as a logistic decay $A^2(T) = \left[1 + exp\left(\frac{T - T^0}{\Delta}\right)\right]^{-s}$, where $T^0$, $\Delta$, and $s$ are fitting parameters. Fitting to experimental thermal conductivity data results in $A_U = 1.316 \times 10^5\,s^{-1}K^{-1}$, $A_{imp} = 4.758 \times 10^6\,s^{-1}$, $a_{CDW} = 5.327 \times 10^8\,s^{-1}\,\text{Å}^2$, $\Theta_U = 120K$, $T^0 = 196.88K$, $\Delta = 3.95K$, and $s \sim 0.122$. The fitted curve is shown in **Fig. 2**. This formulation captures the measured V-shaped behavior of $\kappa(T)$, since CDW scattering rates grow with $w^2A^2$ at intermediate temperatures ($122K \sim 210K$), driving a minimum in thermal conductivity, and diminish once the amplitude collapses, allowing $\kappa(T)$ to recover at elevated temperatures.

The thermal conductivities we measure by TTG are systematically higher than those from the earlier steady-state measurement [49], a difference that reflects both an improved measurement methodology and better sample quality. TTG is a non-contact, all-optical technique: the thermal length scale is set by the optically defined grating period, the diffusivity is read directly from a clean single-exponential decay, and no transducer or wired contact with the sample is required. It is therefore free of the contact thermal resistance, radiative losses, and parasitic heat leaks that limit the accuracy of steady-state methods, yielding a more reliable measure of the intrinsic conductivity. Sample quality reinforces the difference, because impurity concentration strongly governs CDW pinning and hence phonon scattering, as established in related materials such as 2H-$NbS_2$ [61]: impurity pinning stabilizes local CDW regions and enhances phonon scattering, suppressing the conductivity. The pristine single crystals used here, selected by single-crystal X-ray screening for minimal stacking disorder (SI), therefore expose the intrinsic transport that both parasitic losses and impurity pinning otherwise obscure. This combined improvement accounts for the wide scatter in previously reported properties [23,62,63] and confirms that the anomalous reversal is an intrinsic property of clean 2H-$TaSe_2$, resolved by an accurate, non-contact measurement rather than an artifact of disorder or instrumentation. Anomalous thermal conductivity is itself familiar in charge-density-wave systems. Quasi-one-dimensional conductors such as $K_{0.3}MoO_3$, $(TaSe_4)_2I$, and $(NbSe_4)_{10}I_3$ show a weak conductivity minimum, a few percent in size, near the Peierls transition, where phasons of the established CDW carry heat[64]; 1T-$TaS_2$ shows a suppressed, nearly temperature-independent conductivity set by electron-phonon coupling [65]; and an early study read short-range order into the gently rising conductivity of 2H-$TaSe_2$ itself, without resolving a minimum (50). The reversal reported here departs from all of these: its minimum lies roughly 90 K above the ordering transition, deep in the nominally disordered phase; it is large rather than a few-percent correction; and it is driven not by the modes of an established CDW but by phonons scattering off charge order that merely fluctuates, a single mechanism that reproduces the entire curve quantitatively from the measured spatial correlations. Heat conduction is thus governed by order in a phase that nominally has none. Because this control operates through fluctuations rather than a rigid ordered state, it is also more readily tuned: the fluctuating order, and with it the heat flow, responds sensitively to temperature and, by extension, to pressure, strain, and doping, in a way the locked order of conventional CDW conductors does not.

Our results show that the near-universal decrease of thermal conductivity with temperature can fail outright when a fluctuating electronic order, rather than anharmonicity alone, governs phonon scattering. We expect this to be more than a peculiarity of 2H-$TaSe_2$: wherever a charge-neutral order fluctuates over a wide temperature range, it may reshape heat conduction in similar ways, and phase diagrams defined solely by the loss of long-range order may underestimate the temperature window over which such correlations remain active. Most directly, the fluctuating order acts as a control knob for heat flow, switching a phonon-scattering channel on and off. Because this order responds to temperature, strain, electrostatic gating, and controlled disorder, each offers a practical handle to tune heat conduction on demand, enabling thermal switches and related elements that route heat through a material's electronic order. The same physics connects naturally to other correlated systems, most notably the cuprate high-temperature superconductors, in which short-range charge order and fluctuating correlations pervade the pseudogap regime far above the superconducting transition. More broadly, our work shows how the interplay of lattice distortions, electronic instabilities, and defect-mediated heterogeneity can govern thermal transport in low-dimensional quantum materials, and it offers thermal conductivity as a sensitive, nondestructive route to the fluctuating, charge-neutral order that eludes other probes.

## Data availability

The electron microscopy datasets in this study are available through PARADIM [66], a National Science Foundation Materials Innovation Platform. All other raw data supporting the findings of this study are available upon request.

## Code availability

First-principles calculations used the publicly available VASP and TDEP packages, and no custom code was developed for this study. The analysis scripts used to fit the transient thermal grating decays and to evaluate the phonon-CDW scattering model are available from the corresponding author on reasonable request.

**Acknowledgements**

HH, JD, and ZT thank Allan MacDonald for fruitful discussions and suggestions. HH, JD, JC, GA, and ZT thank Jacob Ruff for assistance with X-ray diffraction measurements and helpful discussions. DM and NS thank Saif Siddique for assistance with the preparation of exfoliated samples for TEM analysis. This research used resources of the Advanced Photon Source, a U.S. Department of Energy (DOE) Office of Science user facility operated for the DOE Office of Science by Argonne National Laboratory under Contract No. DE-AC02-06CH11357. Research conducted at the Center for High-Energy X-ray Sciences (CHEXS) was supported by the National Science Foundation (BIO, ENG and MPS Directorates) under awards DMR-1829070 and DMR-2342336. This research made use of the electron microscopy facility of the Platform for the Accelerated Realization, Analysis, and Discovery of Interface Materials (PARADIM), which is supported by the National Science Foundation under Cooperative Agreement No. DMR-2039380 and the Cornell Center for Materials Research shared instrumentation facility. The Thermo Fisher Spectra 300 X-CFEG was acquired with support from PARADIM, an NSF MIP (DMR-2039380), and Cornell University. CS thanks the NSF under 1726077 for the single-crystal X-ray diffraction experiments.

**Funding**

Office of Naval Research award N00014-22-1-2357 (HH, JD, ZT)

National Science Foundation awards DMR-1829070 supporting the Center for High-Energy X-ray Sciences (CHEXS)

National Science Foundation awards DMR-2342336 supporting the Center for High-Energy X-ray Sciences (CHEXS)

National Science Foundation Cooperative Agreement DMR-2039380 supporting the Platform for the Accelerated Realization, Analysis, and Discovery of Interface Materials (PARADIM)

National Science Foundation grant 1726077 supporting single-crystal X-ray diffraction experiments (CS)

U.S. Department of Energy Office of Science user facility support for the APS under Contract No. DE-AC02-06CH11357

**Author contributions**

HH, JD and ZT conceived the study. HH, JD, DMa, AA, DMu and ZT developed the methodology. HH, JD, JC, JK, SK, DMa, NS, GA, CL, CS, MM, BA and AA carried out the experiments and calculations. HH, JD, JC, DMa, NS, CS, MM and CL curated the data, and HH, JD, JC and MM performed the formal analysis. HH, JD, JC, MM and CS prepared the figures. ZT, DN and RC provided materials and resources. CS, DMu and ZT acquired funding. HH, JD and ZT administered the project, and ZT supervised the work. HH, JD and ZT wrote the original draft, and HH, JD, JC, DMa, NS, CS, AA, RC and DMu reviewed and edited the manuscript.

**Competing interests**

The authors declare no competing interests.

**Supplementary Materials**

Materials and Methods

Supplementary Text

Figs. S1 to S7

Tables S1 to S2

References (*62–64*)

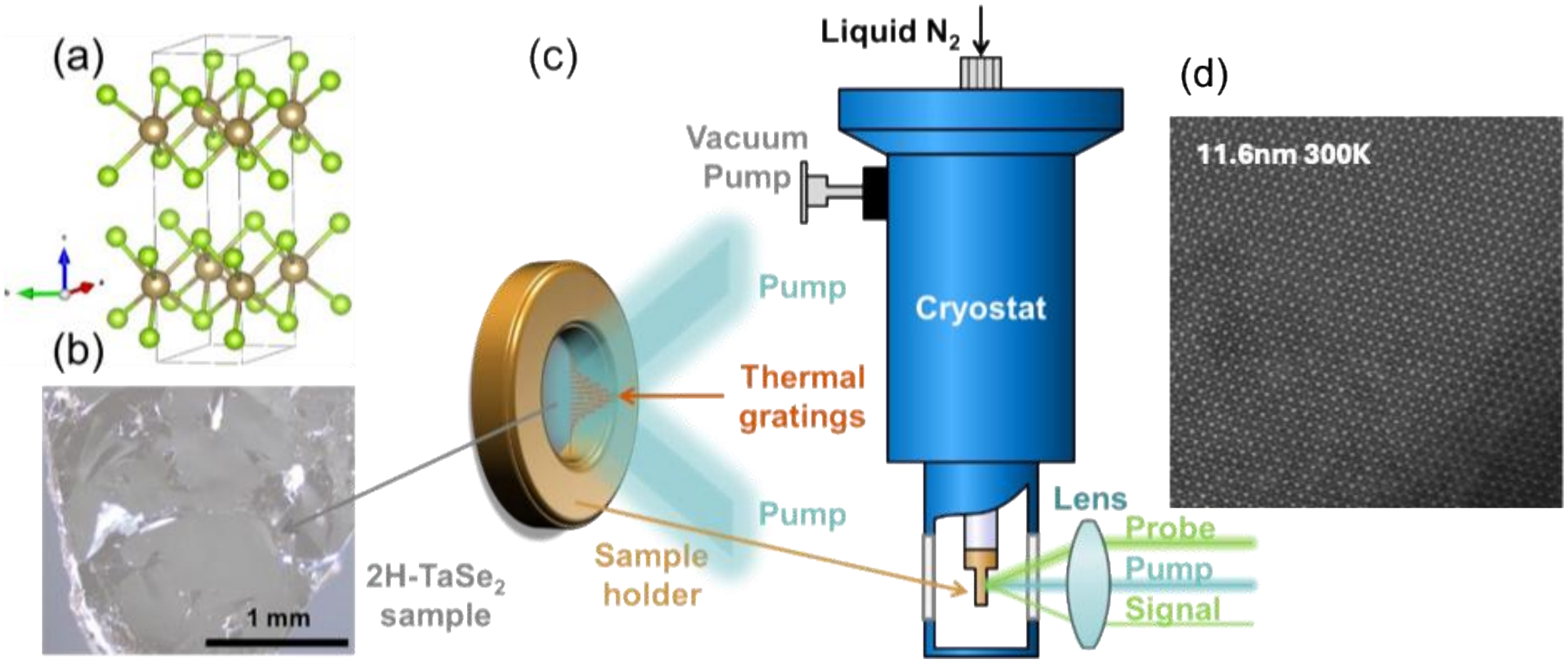

Fig. 1. (a) The unit cell structure of 2H-TaSe2; (b) The microscope photo of an exfoliated 2H-TaSe2 sample ready to be put inside a cryostat and measured. The scale bar is 1 mm; (c) The illustration of a laser-based transient thermal grating (TTG) setup with the sample mounted inside the cryostat. The exfoliated 2H-TaSe2 sample in (b) was mounted on a copper holder, which was loaded into the cryostat and placed under vacuum. Liquid N2 and a heater are used to stabilize the sample temperature via a PID controller. Two pulsed pump lasers are crossed at the sample surface, and a sinusoidal temperature distribution is formed at the crossed region. The continuous wave probe laser detects the decay of the sinusoidal temperature distribution in a reflection geometry of TTG. (d) Atomic-resolution high-angle annular dark-field scanning transmission electron microscopy (HAADF-STEM) image showing the atomically resolved 2H-TaSe2 lattice of a high-quality single crystal. Scale bar, 2 nm.

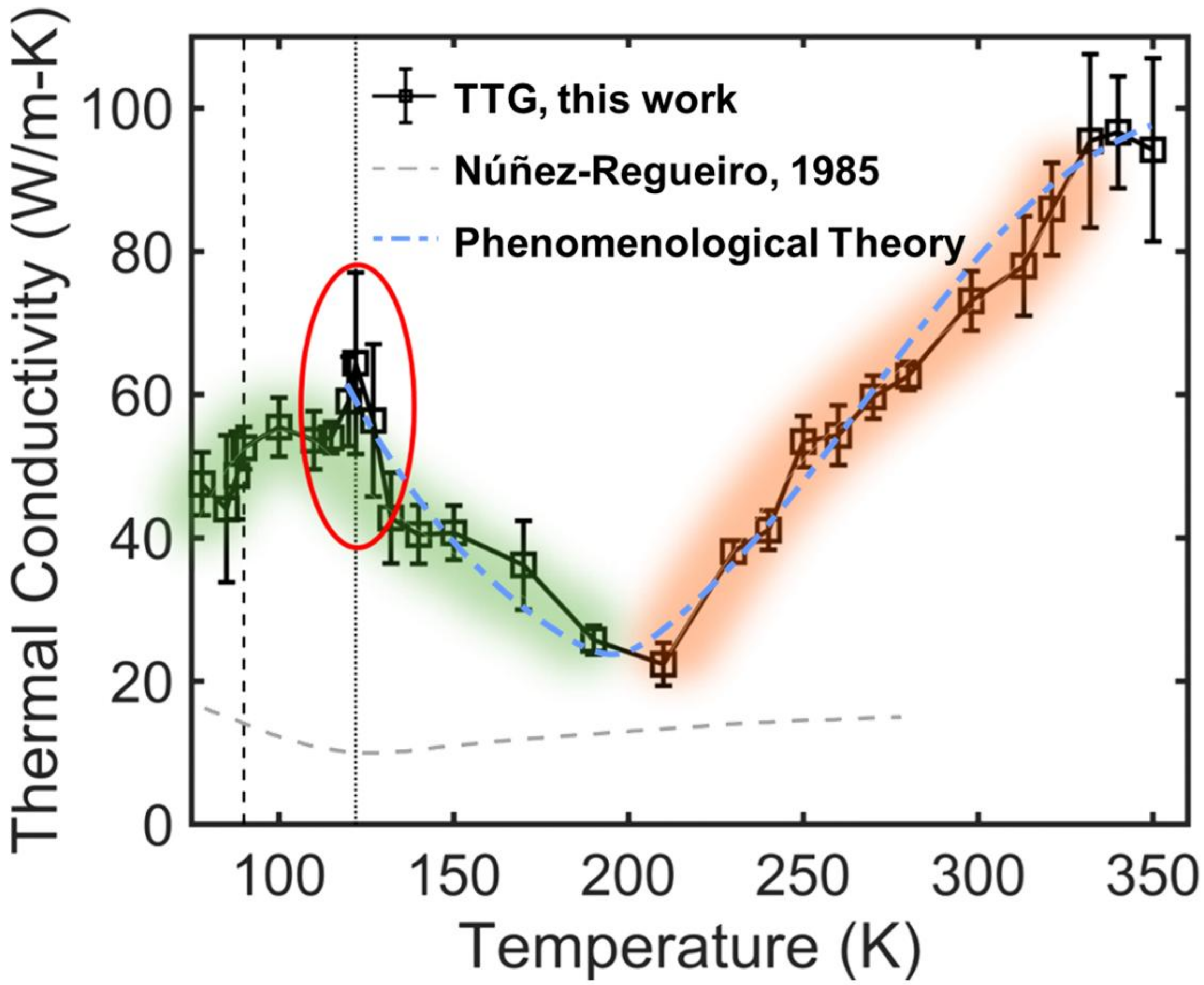

**Fig. 2.** Anomalous nonmonotonic thermal conductivity of 2H-$TaSe_2$ from 78 K to 350 K. The in-plane thermal conductivity peaks sharply at the 122 K transition, falls to a deep minimum near 210 K, and then anomalously reverses, rising with temperature to the highest temperatures measured. The two CDW transitions are marked by a vertical dashed line (90 K) and a dotted line (122 K). The red circle highlights the transition peak, which arises from a λ-shaped specific-heat anomaly that reflects critical-like fluctuations of the CDW order, characteristic of a continuous transition and consistent with a weak first-order crossover. The green and orange shades mark the two regimes flanking the 210 K minimum of the V-shaped trend. The blue dot-dashed curve is the fit from our phenomenological phonon–CDW scattering model. Steady-state data from previous work [49] are plotted for comparison.

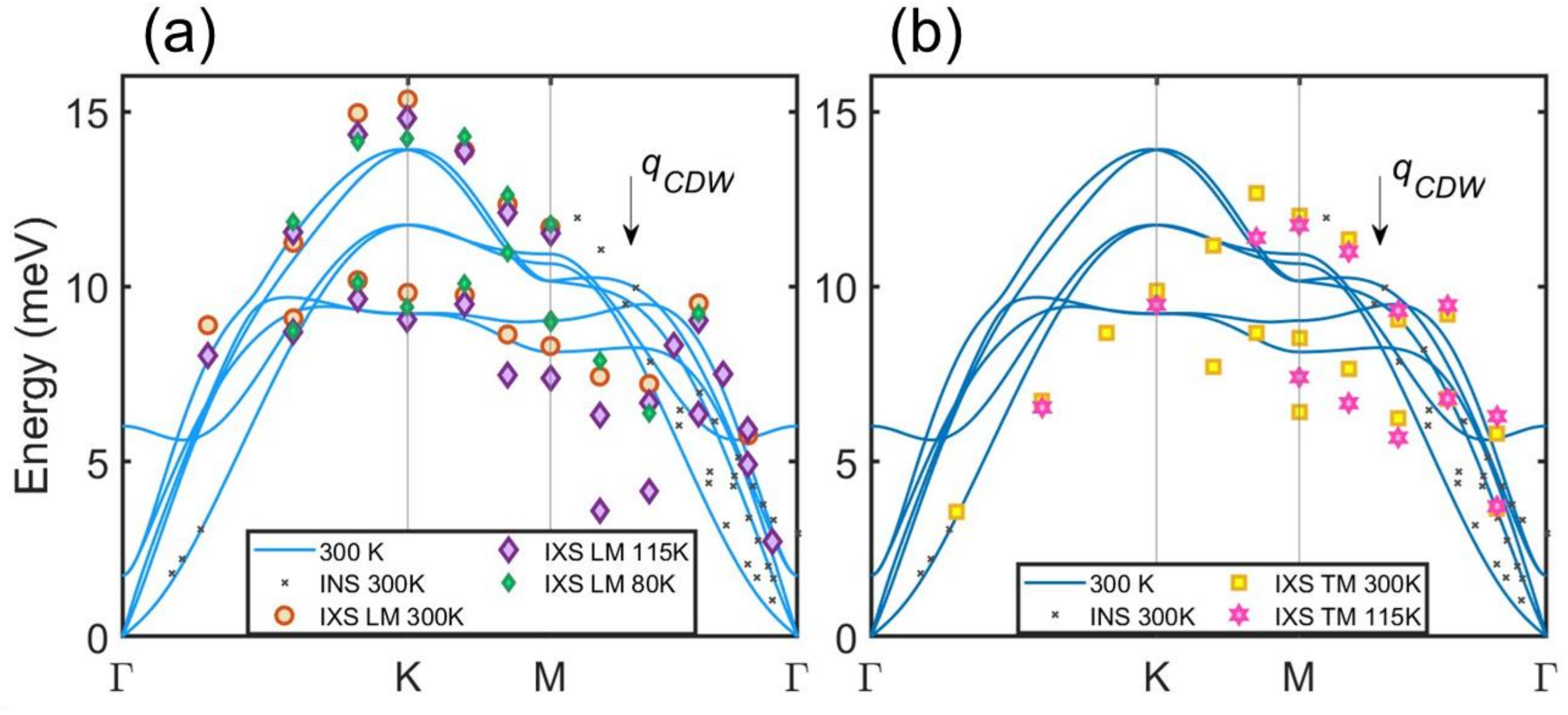

**Fig. 3. Phonon dispersion of 2H-TaSe₂**: the blue lines are from TDEP first principles calculation; the black dots are the room temperature inelastic neutron scattering (INS) measurement from literature [6]. For (a), the red circles are IXS measured longitudinal phonon modes at room temperature; the purple diamonds are IXS measured longitudinal phonon modes at 115 K; the purple diamonds are IXS measured longitudinal phonon modes at 80 K. For (b), the yellow squares are IXS measured transverse phonon modes at room temperature; the pink stars are IXS measured transverse phonon modes at 115 K.

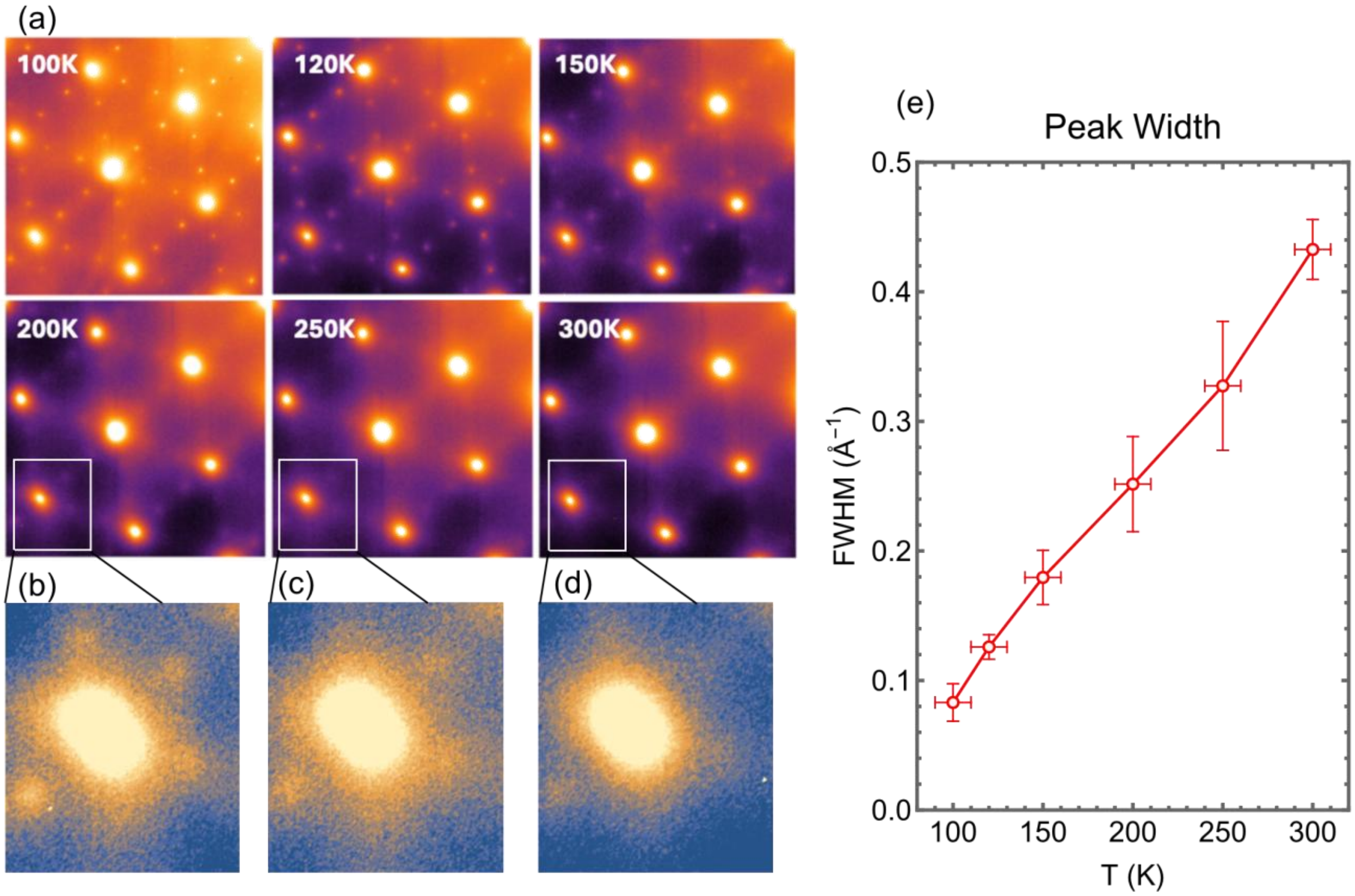

Fig. 4 (a) Temperature-dependent electron diffraction showing Bragg peaks and CDW superlattice at 100K, 120K, 150K, 200K, 250K, and 300K; (b)-(d) zoom-in figure of filtered-out residue peaks at 200K, 250K, and 300K, indicating residue local CDW order at high temperatures; (e) superlattice peak widths from electron diffraction, which increase as temperature increases, indicating diminishing local CDW order. The horizontal temperature error bar comes from the spatial separation between the holder-mounted temperature sensor and the sample position, which introduces uncertainty in the actual sample temperature for each SAED dataset.

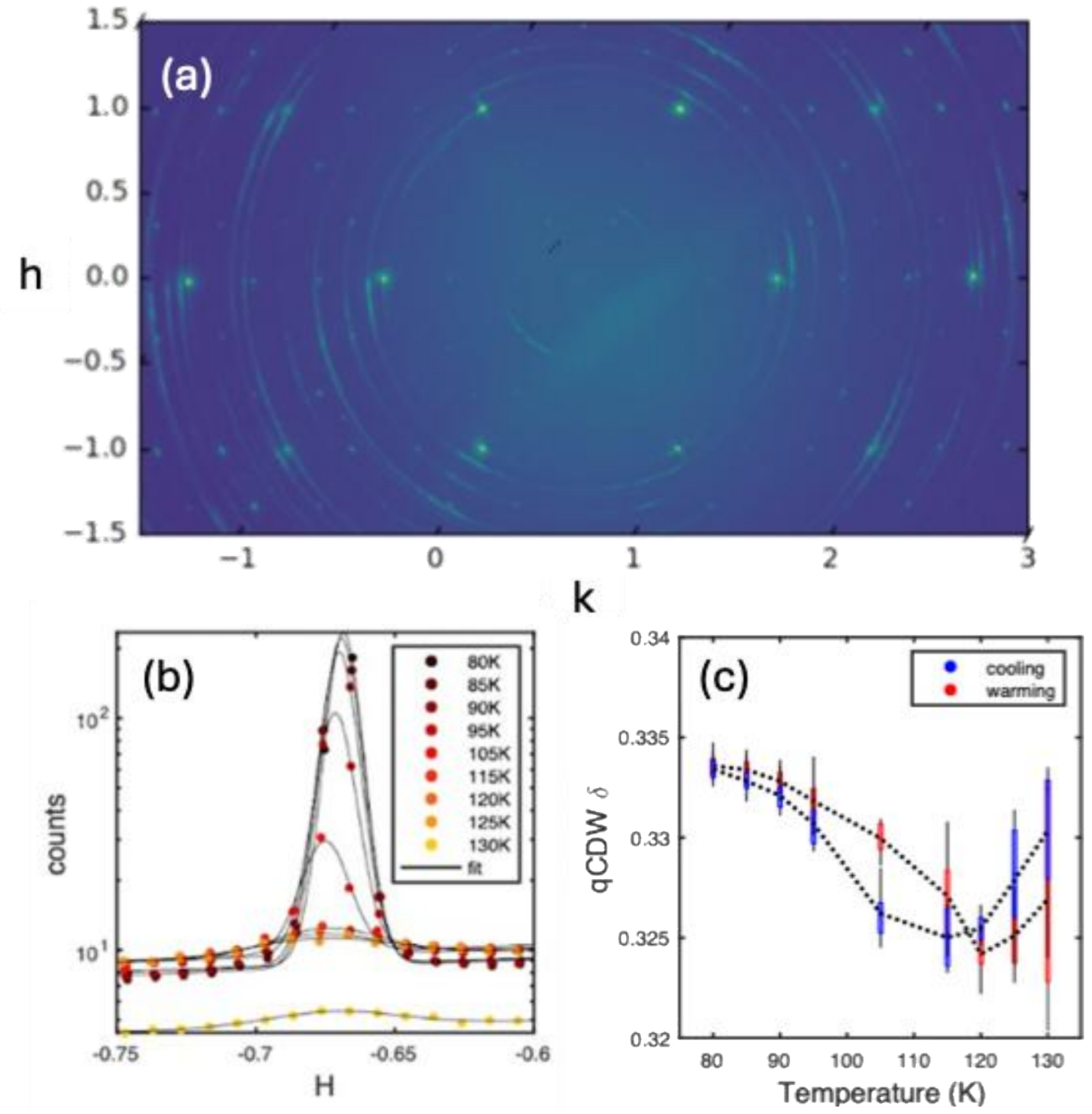

Fig. 5 (a) X-ray scattering data taken at 80 K of the hk0 plane, clean diffraction peaks indicated good crystalline order in the hexagonal plane; (b) Cooling data collected at the (-1/3 - $\delta$, 0, 0) superlattice reflection related to the (001) Bragg peak, where representative line scans at each temperature were fitted with Gaussian profiles to extract the peak position and integrated intensity, and each reported peak position was obtained by averaging over more than 30 symmetry-equivalent superlattice peaks within the same diffraction pattern, with error bars representing the statistical uncertainty of this averaging; (c) Thermal hysteresis of $|\mathbf{q}_{\mathrm{CDW}}|$, indicating latent heat associated with first-order transition.

# Supplementary Materials for

## Anomalous Thermal Conductivity Reversal Driven by a Fluctuating Charge Density Wave

Han Huang[1, †], Jinghang Dai[1, †], Joyce Christiansen-Salameh[1], Jiyoung Kim[1], Samuel Kielar[1], Desheng Ma[2], Noah Schinitzer[3,4], Danrui Ni[5], Gustavo Alvarez[1], Chen Li[1], Carla Slebodnick[6], Mario Medina[1], Bilal Azhar[1], Ahmet Alatas[7], Robert Cava[6], David Muller[2,4], Zhiting Tian[1,*]

**Affiliations:**

[1]Sibley School of Mechanical and Aerospace Engineering, Cornell University, Ithaca, New York 14853, USA.

[2]School of Applied and Engineering Physics, Cornell University, Ithaca, NY 14853, USA.

[3]Department of Materials Science and Engineering, Cornell University, Ithaca, NY 14853, USA.

[4]Kavli Institute at Cornell for Nanoscale Science, Cornell University, Ithaca, NY 14853, USA.

[5]Department of Chemistry, Princeton University, Princeton, New Jersey 08544, USA.

[6]Department of Chemistry, Virginia Tech, Blacksburg, Virginia 24061, USA.

[7]Advanced Photon Source, Argonne National Laboratory, Argonne, Illinois 60439, USA.

*Corresponding author: zhiting@cornell.edu

**Materials and Methods**

Materials

$TaSe_2$ adopts a layered structure belonging to the trigonal or hexagonal space groups, with in-plane lattice constants $a = b \approx 3.45$ Å and a c-axis dimension $c = n \times 6.33$ Å, where 6.33 Å is the single-layer thickness and $n$ is the number of layers per unit cell [67]. At room temperature, four polymorphs exist, all built from edge-sharing $TaSe_6$ trigonal prisms (**Table S1**; **Fig. S1**). The 2H polymorph ($n = 2$, space group $P6_3/mmc$) with $c \approx 12.7$ Å is the thermodynamically stable slow-cooled phase and the subject of the present study. A 3R phase (n = 3, rhombohedral, $c \approx 19.2$ Å) is metastable, and two distinct $n = 4$ variants with $c \approx 25.2$ Å also exist. Above $\sim 800°C$ the coordination switches to octahedral; rapid quenching can trap these high-temperature phases. Because the interlayer stacking order is easily disrupted, broad diffraction peaks along c* and ambiguous c-axis lattice parameters are commonly observed and serve as a diagnostic for stacking faults.

*Crystal Screening Summary*

Single crystals of 2H-$TaSe_2$ were prepared by chemical vapor transport. Thin plate-like crystals were selected from the bulk batch; three candidates (xtl1, xtl2, xtl3) covering a range of lateral dimensions were characterized by single-crystal X-ray diffraction (SCXRD) at the Advanced Photon Source (beam diameter $\sim 120\ \mu m$ FWHM). Each crystal was screened at two to three positions to assess phase purity, in-plane ordering, and stacking coherence along the c-axis. Crystal dimensions and a summary comparison are given in **Table S2**.

All three crystals share a common signature: diffraction is clean along a* and b*, confirming well-ordered individual layers, while diffraction along c* shows varying degrees of peak broadening and off-lattice intensity that indicate stacking faults in the layer packing sequence. The 2H polymorph is nonetheless dominant in every crystal. Among the three, xtl3 shows the highest fraction of indexed reflections (55–74%) and the cleanest c*-axis diffraction, and was selected as the primary sample; xtl2 was mounted as a backup. This screening ensures that the transport reported here reflects high-quality, predominantly 2H material rather than stacking-disordered or impurity-pinned regions.

*Individual Crystal Analysis*

Crystal xtl1. This crystal was screened at two positions but was not mounted for measurement. Both positions show clean a*/b* diffraction (peaks falling on lattice vertices), confirming ordered layers. However, viewing down a*, diffraction peaks along c* are poorly defined with significant inter-peak intensity, indicating pronounced stacking disorder. The fraction of reflections indexed to the 2H cell was only 40–60% (**Fig. S2**). The unit cell refined to $a \approx 3.43 - 3.44$ Å, $c \approx 12.70 - 12.72$ Å ($V \approx 130$ Å$^3$), consistent with the 2H phase.

Crystal xtl2. Screened at three positions. In-plane a*/b* diffraction is clean with only trace off-vertex intensity, suggesting a minor secondary component. The c*-axis histogram shows small bumps between the expected 2H ($c \approx 12.7$ Å) peak positions, consistent with a small admixture of an $n = 4$ phase ($c \approx 25.2$ Å). The fraction indexed to 2H ranges from 35 to 53% (**Fig. S3**). Although individual positions of xtl1 sometimes appear cleaner, xtl2 is more spatially uniform; it was therefore mounted as a backup sample.

Crystal xtl3 (primary sample). Screened at three positions. The c*-axis diffraction is markedly cleaner than for xtl1 and xtl2: the peak-spread histograms show well-defined maxima coinciding with the expected 2H periodicity, with little inter-peak noise. Viewing down a*, peaks along c* fall cleanly on the lattice vertices. The fraction of reflections indexed to the 2H cell ranges from 56 to 74% (**Fig. S4**), substantially higher than for the other crystals. The unit cell parameters are consistent across all three positions: $a \approx 3.44$ Å, $c \approx 12.69$ Å ($V \approx 130$ Å$^3$), in excellent agreement with literature values for 2H-$TaSe_2$. On the basis of these criteria, xtl3 was selected as the primary sample.

Methods

*Transient Thermal Grating*

Transient thermal grating (TTG) spectroscopy is a non-contact, laser-based pump-probe technique [41–46]. As schematically illustrated in **Fig. S5**, a pulsed pump laser is split by a transmission diffraction grating (phase mask) into two first-order beams, which are subsequently collimated and focused by a pair of achromatic doublets (AC1 and AC2) onto the sample surface, where they intersect at an angle $\theta$. The resulting optical interference pattern deposits energy with a spatially sinusoidal profile of period

$$\Lambda = \frac{\lambda_{\mathrm{p}}}{2\sin\left(\frac{\theta}{2}\right)} = \frac{2\pi}{q} \qquad (S1)$$

where $\lambda_p$ is the pump wavelength and $q = 2\pi/\Lambda$ is the grating wavevector. A continuous-wave probe beam co-incident on the same region diffracts off the transient grating and is collected by a photodetector. Heterodyne detection, achieved by superimposing a reference beam derived from the same probe source, is employed to separate the in-phase (amplitude grating) and quadrature (phase grating) components of the diffracted signal, substantially improving the signal-to-noise ratio.

Absorption of the pump pulse deposits energy in a spatially periodic pattern along $x$, instantly establishing a sinusoidal temperature field on the sample surface. Two simultaneous physical responses are excited. First, the spatially modulated temperature drives thermal diffusion: the temperature grating decays as heat flows from hot to cold regions, erasing the periodic pattern on a timescale set by the in-plane thermal diffusivity $\alpha_x$. This is described by the thermal diffusion equation

$$\rho c_{\mathrm{v}}\frac{\partial T}{\partial t} - k_{\mathrm{x}}\frac{\partial^2 T}{\partial x^2} - k_{\mathrm{r}}\frac{\partial^2 T}{\partial z^2} = Q(x,z,t) \quad (S2)$$

where $\rho$ is density, $c_{\mathrm{v}}$ is specific heat capacity, $k_{\mathrm{x}}$ and $k_{\mathrm{r}}$ are the in-plane and out-of-plane thermal conductivity components, and $Q(x,z,t)$ is the absorbed heat density. For a delta-function pump pulse with a cosine intensity profile, the surface temperature evolves as

$$T(z=0,x,t)=\frac{A}{\sqrt{\alpha_r t}}\cos(qx)\exp(-\alpha_x q^2 t) \quad (S3)$$

where $A$ is an amplitude and $\alpha_i = k_i/\rho c_\mathrm{v}$ are the directional thermal diffusivities. The exponential factor $\exp(-\alpha_x t)$ governs the decay of the thermal grating with a characteristic relaxation time $\tau = 1/(\alpha_x q^2)$.

Second, the spatially periodic thermal expansion simultaneously launches surface acoustic waves (SAWs): counter-propagating Rayleigh-like waves form a standing-wave grating whose frequency $f = v_p/\Lambda$ is set by the SAW velocity $v_p$ and the grating period. The SAW oscillations are superimposed on the thermal decay in the detected signal and carry information about the elastic properties of the material.

In the amplitude grating configuration (heterodyne phases $\Phi_1 = 0, \Phi_2 = \pi$), the detected signal is proportional to the real part of the reflectivity change and follows

$$I(t)=\frac{A}{\sqrt{t}}\exp(-q^2\alpha t)+ B \quad (S4)$$

where $A$ and $B$ are fitting constants and $\alpha = \alpha_x$ is the in-plane thermal diffusivity. The thermal diffusivity is extracted by nonlinear least-squares fitting of the measured amplitude grating decay to $f(x) = a\exp(-bx) + d$, where fitting constants are simplified to $a$, $b$, and $d$. An example fit at room temperature is shown in **Fig. S6**. The thermal conductivity is then obtained as $k = \alpha\rho c_v$.

TTG measurements were performed in reflection geometry, appropriate for opaque crystalline samples. The 2H-$TaSe_2$ crystal surfaces were mirror-polished to a roughness below the probe wavelength to minimize diffuse scattering. The grating excitation area (beam diameter ~200–500 μm) was kept large relative to $\Lambda$ to reduce edge effects and limit the peak temperature rise to $\Delta T < 10\ K$, well within the linear elastic and linear thermal response regime. No metallic transducer layer was required, in contrast to TDTR/FDTR, as 2H-$TaSe_2$ absorbs directly at the pump wavelength.

*Specific Heat Extrapolation*

The transient thermal grating (TTG) technique is employed to measure thermal diffusivity ($\alpha$). To determine the thermal conductivity ($\kappa$), the measured diffusivity is multiplied by the density ($\rho$) and specific heat capacity ($C_p$), governed by the relationship

$$\kappa = \alpha \rho C_p \qquad (S5)$$

The specific heat capacity values for 2H-$TaSe_2$ were adapted from literature [47]. As the available experimental data is limited to 160 K, the specific heat at higher temperatures was extrapolated using the Debye model

$$C_p(T) = 9Nk_B \left(\frac{T}{\Theta_D}\right)^3 \int_0^{\frac{\Theta_D}{T}} \frac{x^4 e^x}{(e^x - 1)^2} dx + \left(C_p - C_v\right) \qquad (S6)$$

where $N$ is the number of atoms per unit volume (or formula unit), $k_B$ is the Boltzmann constant, $\Theta_D$ is the Debye temperature, and $x = \frac{\hbar\omega}{k_B T}$ is a dimensionless integration variable. For the temperature range of interest ($T < 350K$) , we can regard the thermal expansion part $\left(C_p - C_v\right)$ as negligible. Our fitted parameters are $Nk_B = 24.4 \frac{mJ}{mol \cdot K}$ and $\Theta_D = 262K$, which reproduces the experimental data from [47] with accuracy sufficient for extrapolation purposes (**Fig. S7**).

*Temperature Dependent Effective Potentials Method for Phonon Dispersion*

First-principles calculations were performed using the temperature-dependent effective potential (TDEP) method [37,38] to obtain finite-temperature lattice dynamics for 2H-$TaSe_2$. All density functional theory (DFT) calculations, including ab initio molecular dynamics (AIMD) simulations, were carried out using the [39,40] with the Perdew-Burke-Ernzerhof (PBE) [68] exchange-correlation functional within the generalized gradient approximation (GGA).

The initial crystal structure of 2H-$TaSe_2$ was obtained from the Materials Project database. An ionic relaxation was performed using the conjugate gradient algorithm (IBRION = 2) with only the cell volume allowed to change (ISIF = 7) to determine optimal lattice parameters while preserving the hexagonal crystal symmetry. The relaxed lattice parameters ($a = 3.39$ Å, $c = 12.60$ Å) are in good agreement with experimental values ($a = 3.32$ Å, $c = 12.07$ Å) and previous computational studies ($a = 3.34$ Å, $c = 12.54$ Å).

A 4×4×1 supercell containing 96 atoms was constructed from the relaxed unit cell using the TDEP software. An initial atomic configuration at 300 K was generated using TDEP based on the reported Debye temperature of 202 K. Gamma-point AIMD simulations (IBRION = 0) were then performed using an NVT ensemble with a Nosé-Hoover thermostat. A time step of 2 fs and a plane-wave cutoff energy of 450 eV were used. The system was equilibrated until temperature oscillations decreased to $\pm 10\%$ of the target temperature and external pressure remained below 0.5 GPa, which required approximately 7 ps. Following equilibration, an additional 1 ps AIMD trajectory was collected, with atomic forces recorded at each self-consistent field (SCF) step.

The AIMD trajectory data were processed using the "process_outcar" utility within TDEP to extract position-force pairs. An initial set of second-order and third-order interatomic force constants (IFCs) was obtained by fitting these data using least-squares regression. To improve the accuracy of the IFCs, 80 uncorrelated supercell configurations at 300 K were generated using TDEP based on Maxwell-Boltzmann statistics. The convergence of thermal conductivity with respect to the number of configurations was verified, with 80 configurations found to yield converged results. Static SCF calculations (IBRION = −1) were performed on each configuration using a 3×3×1 Monkhorst-Pack k-point mesh. The forces from these calculations were collected and processed to obtain the final set of temperature-dependent IFCs.

The phonon dispersion relations shown in Fig. 3 of the main text were calculated from the second-order IFCs using TDEP. The calculated acoustic phonon branches are in good agreement with both our IXS measurements and previous inelastic neutron scattering (INS) data from the literature. The in-plane thermal conductivity ( $\kappa_{xx} = \kappa_{yy} \equiv \kappa_{\parallel}$ ) was computed by solving the linearized Boltzmann transport equation (BTE) within TDEP using an integration mesh of 26×26×26. The heat capacity was calculated from the phonon density of states using the temperature-dependent IFCs.

DFT and AIMD calculations were performed on the Summit supercomputer at Oak Ridge National Laboratory, supported by the Office of Science of the U.S. Department of Energy under Contract No. DE-AC05-00OR22725. TDEP post-processing was performed on the Expanse supercomputer through the Extreme Science and Engineering Discovery Environment (XSEDE), supported by NSF grant number ACI-1548562.

**Supplementary Text**

Derivation of Phenomenological Thermal Conductivity Model

*Foundations*

Here we derive the phenomenological model for thermal conductivity in the presence of fluctuating CDW order. The goal is to find the amplitude and linewidth dependence of thermal conductivity under the assumption that CDW fluctuation generates phonon scattering. Starting from the CDW order parameter $\psi(\mathbf{r}) = A(\mathbf{r})e^{i\phi(\mathbf{r})}$, with $A(\mathbf{r})$ the amplitude and $\phi(\mathbf{r})$ the phase. In the presence of spatial deformation, $\phi(\mathbf{r}) = \mathbf{q}_{\mathrm{CDW}} \cdot \mathbf{r} + qu(\mathbf{r})$, where $\mathbf{q}_{\mathrm{CDW}} \cdot \mathbf{r}$ gives the ideal periodic modulation, and $qu(\mathbf{r})$ describes slow, long-wavelength distortions with $q = |\mathbf{q}|$ and $u(\mathbf{r})$ the displacement along the CDW normal direction. Due to the physical fact that a perfectly uniform CDW only produces a periodic renormalization of the equilibrium atom positions (giving rise to Bragg satellites) but not diffuse phonon scattering, we assume that phonon scattering comes from the fluctuation part. Thus, only spatial variations of $\psi$ generate phonon-scattering inhomogeneities. If we write down the free energy density of the system as

$$f = f_{\mathrm{CDW}}[\psi] + f_{\mathrm{elastic}}[\epsilon_{ij}] + f_{\mathrm{int}}[\psi, \epsilon_{ij}], \qquad (S7)$$

where $\epsilon_{ij}$ is the symmetrized strain tensor, the CDW-induced modulations of the elastic moduli should arise from gradients of the order parameter

$$\delta C_{ijkl}(\mathbf{r}) \propto \mathcal{F}_{ijkl}[\nabla\psi(\mathbf{r})] = \mathcal{F}_{ijkl}[\nabla A(\mathbf{r}), A(\mathbf{r})\nabla u(\mathbf{r})]. \qquad (S8)$$

Since $\delta C_{ijkl}(\mathbf{r}) = \frac{\partial f_{\mathrm{int}}}{\partial \epsilon_{ij} \partial \epsilon_{kl}}$ and terms linear in $\nabla\psi$ are generally unforbidden for the coupling term, the leading order of $f_{\mathrm{int}}$ that gives rise to nonzero $\delta C_{ijkl}(\mathbf{r})$ is proportional to $\epsilon_{ij}\epsilon_{kl}\nabla\psi$. Thus, the simplest phenomenological form of $\delta C_{ijkl}(\mathbf{r})$ is

$$\delta C_{ijkl}(\mathbf{r}) \propto \nabla\psi. \qquad (S9)$$

We can now write down the elastic Hamiltonian based on the modulation of elastic moduli. Let $\boldsymbol{\eta}(\mathbf{r}, t)$ denote the lattice displacement field and $\epsilon_{ij}(\mathbf{r}, t) = \frac{1}{2}(\partial_i \eta_j + \partial_j \eta_i)$. The elastic energy of an inhomogeneous medium is

$$E_{\mathrm{el}} = \frac{1}{2}\int d^3r \left[\rho \dot{\boldsymbol{\eta}}^2(\mathbf{r}, t) + C_{ijkl}(\mathbf{r})\epsilon_{ij}(\mathbf{r}, t)\epsilon_{kl}(\mathbf{r}, t)\right] \qquad (S10)$$

where $\rho$ is the mass density and $C_{ijkl}(\mathbf{r}) = C_{ijkl}^{(0)} + \delta C_{ijkl}(\mathbf{r})$ the total position-dependent elastic moduli. The Hamiltonian is then

$$H_{\text{el}} = H_0 + H' \qquad (S11)$$

with

$$H_0 = \frac{1}{2}\int d^3r\left[\rho\dot{\boldsymbol{\eta}}^2 + C_{ijkl}^{(0)}\epsilon_{ij}\epsilon_{kl}\right], \qquad (S12)$$

$$H' = \frac{1}{2}\int d^3r\delta C_{ijkl}(\mathbf{r})\epsilon_{ij}(\mathbf{r},t)\epsilon_{kl}(\mathbf{r},t) \qquad (S13)$$

Here $H_0$ describes the unperturbed phonon modes in a homogeneous crystal, and $H'$ describes scattering from the elastic inhomogeneity. The Hamiltonian can be quantized using the standard phonon notation

$$\boldsymbol{\eta}(\mathbf{r},t) = \sum_{\mathbf{k},s}\sqrt{\frac{\hbar}{2\rho V\omega_{\mathbf{k}s}}}\left[\hat{\mathbf{e}}_{\mathbf{k}s}a_{\mathbf{k}s}e^{i(\mathbf{k}\cdot\mathbf{r}-\omega_{\mathbf{k}s}t)} + \hat{\mathbf{e}}_{\mathbf{k}s}a_{\mathbf{k}s}^{\dagger}e^{-i(\mathbf{k}\cdot\mathbf{r}-\omega_{\mathbf{k}s}t)}\right] \qquad (S14)$$

where $\omega_{\mathbf{k}s}$ is the dispersion of a phonon with momentum $\boldsymbol{k}$ and branch index $s$, $V$ is the crystal volume, $\hat{\mathbf{e}}_{\mathbf{k}s}$ is the polarization vector of, and $a_{\mathbf{k}s}$, $a_{\mathbf{k}s}^{\dagger}$ are phonon annihilation and creation operators. The corresponding strain tensor in Fourier space is thus linear in $\boldsymbol{k}$ and in the phonon operators. The perturbation $H'$ couples two phonon states via the product of two strain fields. Substituting the mode expansion and keeping terms bilinear in $(a^{\dagger}a)$ such that one phonon into another, we obtain schematically

$$H' \simeq \frac{1}{2}\sum_{\mathbf{k},s}\sum_{\mathbf{k}',s'}\mathcal{M}_{ss'}(\mathbf{k},\mathbf{k}')a_{\mathbf{k}'s'}^{\dagger}a_{\mathbf{k}s} + (\text{ other terms }), \qquad (S15)$$

where the matrix element is

$$M_{ss'}(\mathbf{k},\mathbf{k}') \propto \sqrt{\frac{\hbar}{2\rho V\omega_{\mathbf{k}s}}}\sqrt{\frac{\hbar}{2\rho V\omega_{\mathbf{k}'s'}}}\hat{e}_i^{(s)}(\mathbf{k})\hat{e}_j^{(s')}(\mathbf{k}')k_l k_m'\delta C_{ijlm}(\mathbf{q}) \qquad \mathbf{q} = \mathbf{k}' - \mathbf{k}, \qquad (S16)$$

Here $\delta C_{ijlm}(\mathbf{q})$ is the Fourier transform of the elastic inhomogeneity,

$$\delta C_{ijlm}(\mathbf{q}) = \int d^3re^{-i\mathbf{q}\cdot\mathbf{r}}\delta C_{ijlm}(\mathbf{r}). \qquad (S17)$$

The transition matrix element for scattering a phonon $|\boldsymbol{k}, s\rangle$ into $|\boldsymbol{k}', s'\rangle$ is thus

$$M_{\mathbf{k}s\to\mathbf{k}'s'} = \langle \mathbf{k}', s'|H'|\mathbf{k}, s\rangle \propto \hat{e}^{i}_{\mathbf{k},s}\hat{e}^{j}_{\mathbf{k}',s'}k_l k'_m \delta C_{ijlm}(\mathbf{q}). \qquad (S18)$$

and

$$\tau^{-1}_{\mathrm{CDW}}(\mathbf{k}, s) = \sum_{\mathbf{k}',s'} W_{\mathbf{k}s\to\mathbf{k}'s'} = \frac{2\pi}{\hbar}\sum_{\mathbf{k}',s'} |M_{\mathbf{k}s\to\mathbf{k}'s'}|^2 \delta(\omega_{\mathbf{k}s} - \omega_{\mathbf{k}'s'}). \qquad (S19)$$

Since $\delta C_{ijlm}(\mathbf{r})$ is a static random field produced by the fluctuating CDW texture, it is natural to average over disorder domains

$$\langle \tau^{-1}_{\mathrm{CDW}}(\mathbf{k}, s)\rangle \propto \sum_{s'} \int d^3q \langle |\delta C_{ijlm}(\mathbf{q})|^2\rangle G_{ss'}(\mathbf{k}, \mathbf{q})\delta\big(\omega_{\mathbf{k}s} - \omega_{\mathbf{k}+q,s'}\big) \qquad (S20)$$

where $G_{ss'}$ contains smooth products of polarizations and wavevectors from Eq. (S18). Under the assumption that $\delta C_{ijkl}(\mathbf{r}) \propto \nabla\psi$, we can conclude that

$$\tau^{-1}_{\mathrm{CDW}}(T) \propto \langle |\nabla\psi(\mathbf{r})|^2\rangle = \langle |\nabla A|^2\rangle + q^2\langle A^2(\nabla u)^2\rangle. \qquad (S21)$$

*Connection to Diffraction Observables*

To make connections to temperature-dependent SAED observables, i.e. peak intensity $I_{int}(T)$ and linewidth $w(T)$, we adopt the Ornstein–Zernike form for the correlation function of $u(\mathbf{r})$ [69],

$$H_{OZ}[u] = \frac{K}{2}\int d^d r[(\nabla u)^2 + \xi^{-2}u^2] \qquad (S22)$$

with $\xi$ is the coherence length, which gives the propagator

$$\langle |u_{\mathbf{q}}|^2\rangle = \frac{k_B T}{K(q^2 + \xi^{-2})}. \qquad (S23)$$

Therefore, by Parseval's theorem,

$$\langle (\nabla u)^2\rangle = \langle \nabla u(\mathrm{r}) \cdot \nabla u(\mathrm{r})\rangle = \int d^d q q^2 \langle |u_{\mathrm{q}}|^2\rangle = \int d^d q q^2 \frac{k_B T}{q^2 + \xi^{-2}}. \qquad (S24)$$

Introducing a UV cutoff $\Lambda$ and define $\xi^{-1} \equiv \kappa$,

$$\langle(\nabla u)^2\rangle = \frac{k_B T}{(2\pi)^2}\int_0^\Lambda 2\pi q dq \frac{q^2}{q^2+\xi^{-2}} = \frac{k_B T}{2\pi}\int_0^\Lambda dq \frac{q^3}{q^2+\kappa^2}$$

$$= \frac{k_B T}{2\pi}\left[\frac{\Lambda^2}{2} - \frac{\kappa^2}{2}(\ln(\Lambda^2+\kappa^2) - \ln\kappa^2)\right], \qquad (S25)$$

The first term is UV divergent and independent of $\kappa$, representing the short-wavelength (sub-lattice spacing) physics that is not of interest to our CDW physics and we only keep the $\kappa$-dependent part

$$\Delta\langle(\nabla u)^2\rangle = -\frac{k_B T}{4\pi c}\kappa^2[\ln(\Lambda^2+\kappa^2) - \ln\kappa^2]. \qquad (S26)$$

For $\Lambda^2 \gg \kappa^2$ (i.e. large coherence length near the critical temperature), we have the leading order of $\ln(\Lambda^2+\kappa^2) - \ln\kappa^2 \simeq \ln\Lambda^2 + \mathcal{O}\left(\frac{\kappa^2}{\Lambda^2}\right)$. Therefore,

$$\Delta\langle(\nabla u)^2\rangle \propto \xi^{-2} \propto w^2(T) \qquad (S27)$$

On the other hand,

$$I_{\text{int}}(T) \propto \langle|\psi|^2\rangle \approx \langle A^2\rangle \qquad (S28)$$

and

$$\langle|\nabla A|^2\rangle \propto \frac{\langle A^2\rangle}{\xi^2}, \qquad (S29)$$

we can conclude that the temperature dependence of phonon-fluctuating CDW scattering is given by

$$\tau_{\text{CDW}}^{-1}(T) \propto A^2(T)w^2(T). \qquad (S30)$$

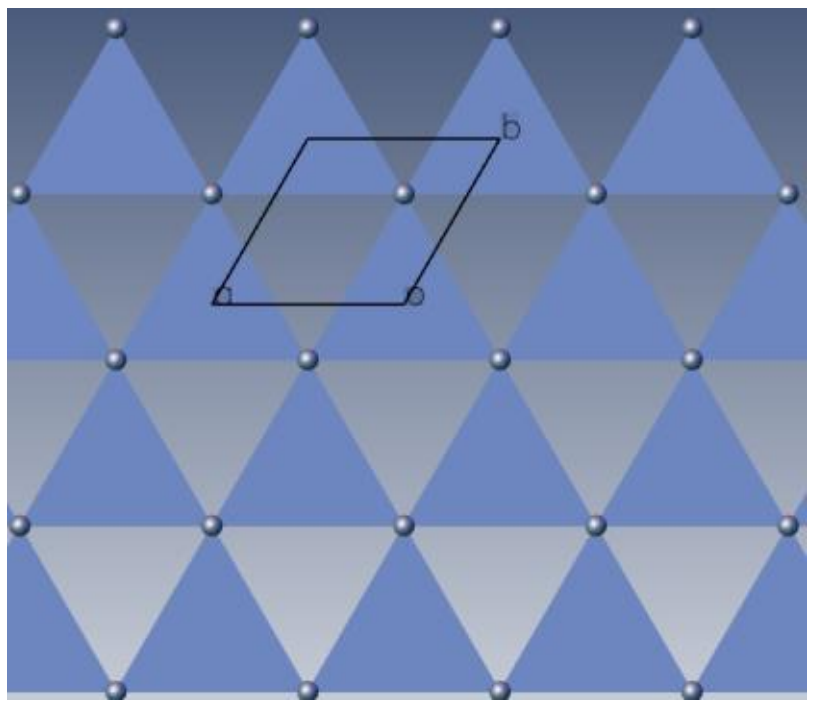

**Fig. S1. View down the c-axis showing the layers of edge-sharing $TaSe_6$ trigonal prisms characteristic of the room-temperature $TaSe_2$ polymorphs.**

| | xtl1 — Position 1 | xtl1 — Position 2 |
|---|---|---|
| Face view | 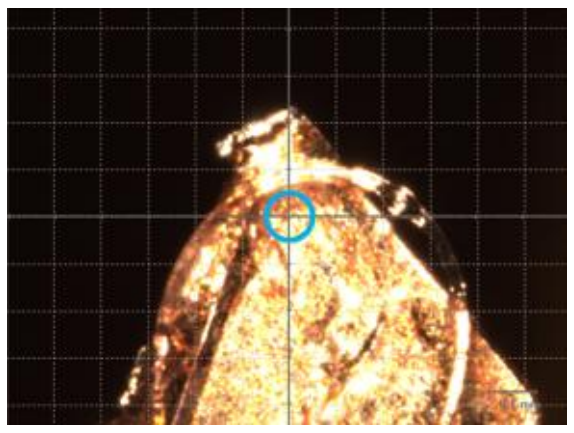 | 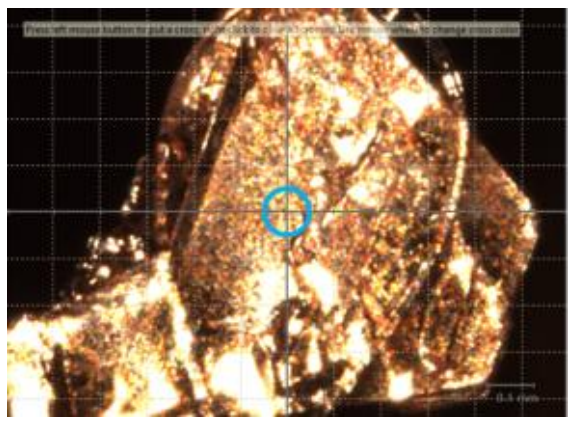 |
| Diff. down c* | 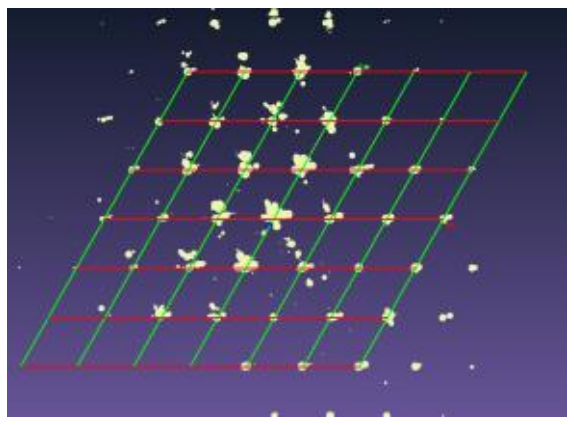 | 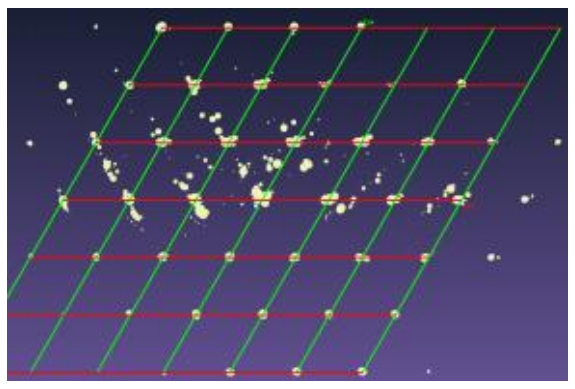 |
| Diff. down a* | 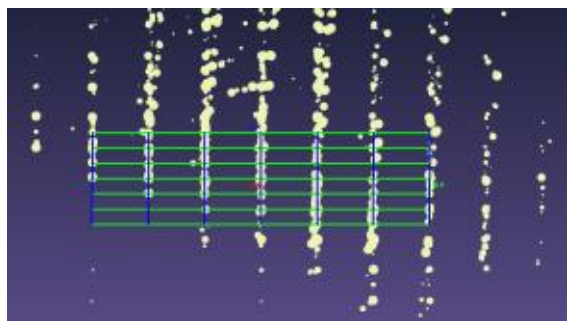 | 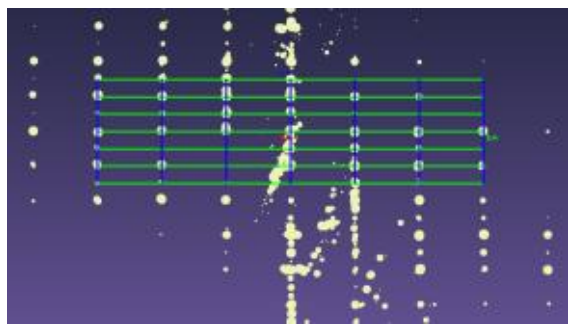 |
| c* histogram | 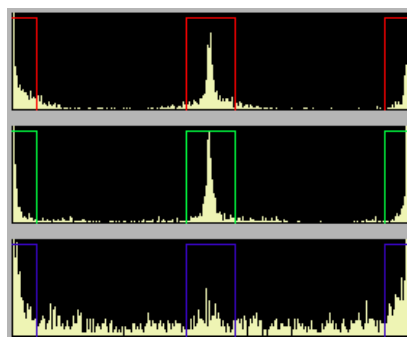 | 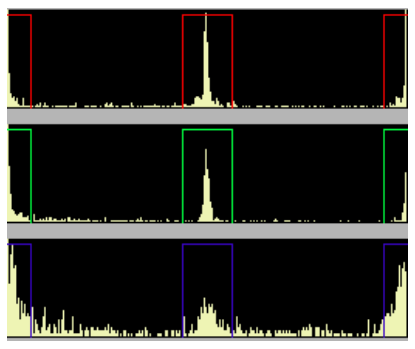 |
| Unit cell (Å) | a = 3.438(3), b = 3.436(2), c = 12.718(4) α = 90.16(4), β = 89.99(5), γ = 119.69(9) V = 130.50(15) $Å^3$ | a = 3.428(3), b = 3.432(3), c = 12.696(2) α = 90.06(3), β = 90.04(3), γ = 119.80(8) V = 129.61(14) $Å^3$ |
| Indexed (%) | 541/1341 (40.3%) | 657/1098 (59.8%) |

**Fig. S2. Single-crystal XRD screening data for xtl1 at two positions.** Rows from top: optical face view (blue circle = 100 $\mu m$ X-ray beam footprint); reconstructed reciprocal-lattice precession image down c*; precession image down a*; peak-spread histogram along a*, b*, c*. The poor definition of peaks along c* and elevated inter-peak histogram counts indicate substantial stacking disorder.

| | xtl2 — Position 1 | xtl2 — Position 2 | xtl2 — Position 3 |
|---|---|---|---|
| Face view | 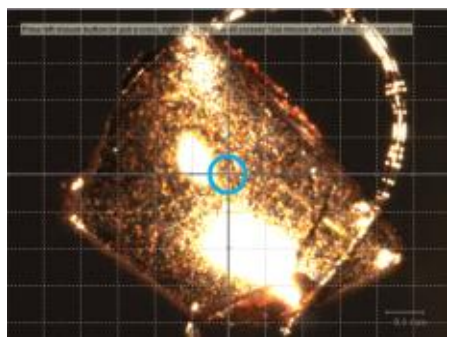 | 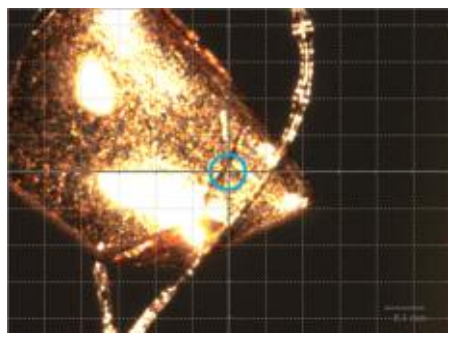 | 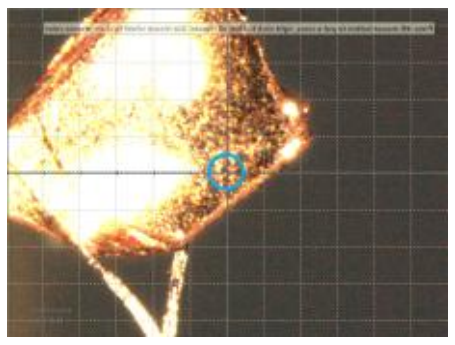 |
| Diff. down c* | 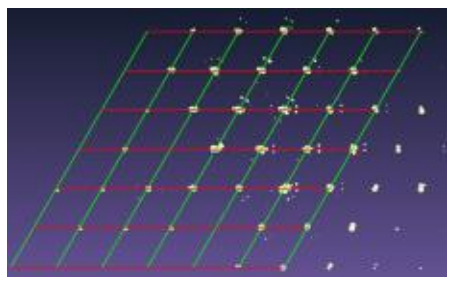 | 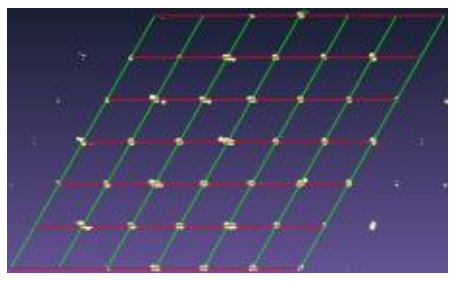 | 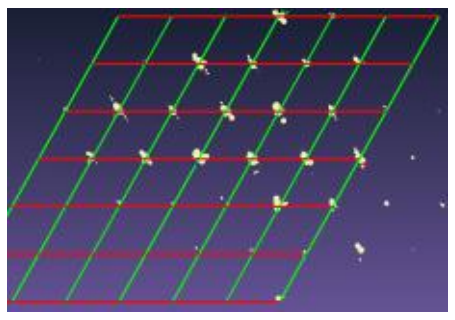 |
| Diff. down a* | 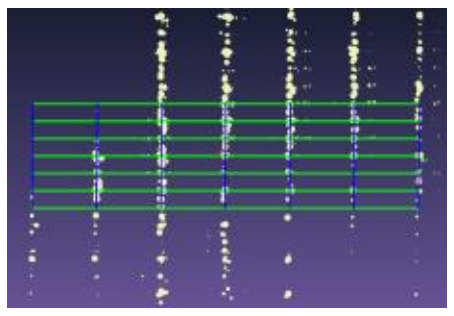 | 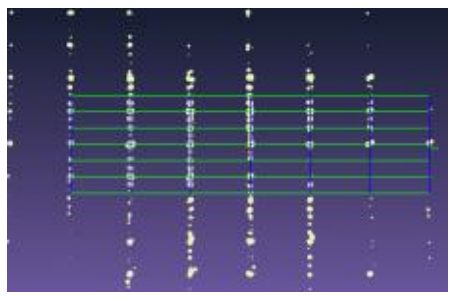 | 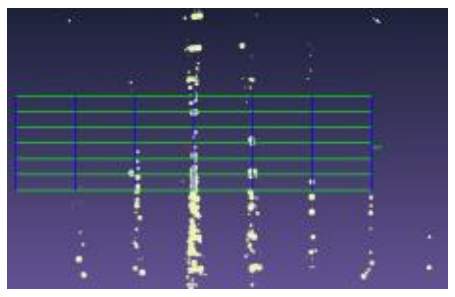 |
| c* histogram | 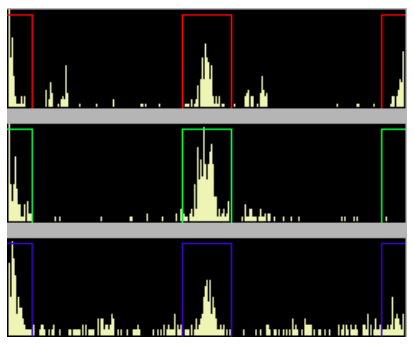 | 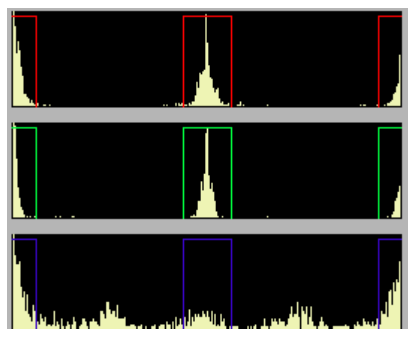 | 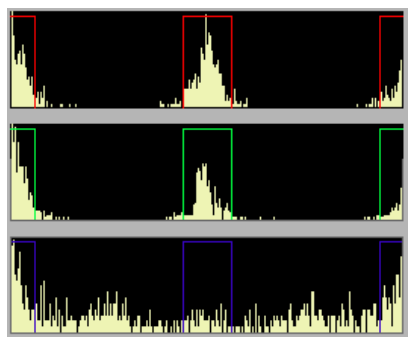 |
| Unit cell (Å) | a=3.437, b=3.442 c=12.683 V=129.97 Å³ | a=3.438, b=3.445 c=12.572 V=129.12 Å³ | a=3.431, b=3.370 c=12.771 V=128.8 Å³ |
| Indexed (%) | 519/1359 (38.2%) | 576/1083 (53.2%) | 269/758 (35.5%) |

**Fig. S3. Single-crystal XRD screening data for xtl2 at three positions.** Row labels as in Fig. S2. Small off-vertex peaks along a*/b* and minor bumps in the c* histogram indicate a trace $n = 4$ impurity phase. xtl2 was mounted as a backup sample due to its greater spatial uniformity.

| | xtl3 — Position 1 | xtl3 — Position 2 | xtl3 — Position 3 |
|---|---|---|---|
| Face view | 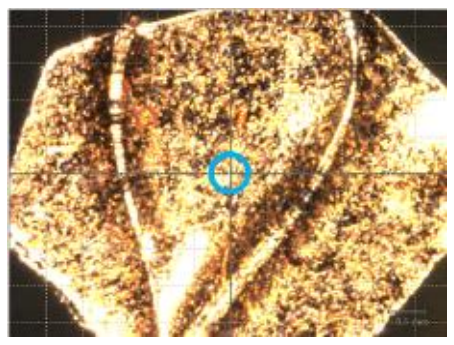 | 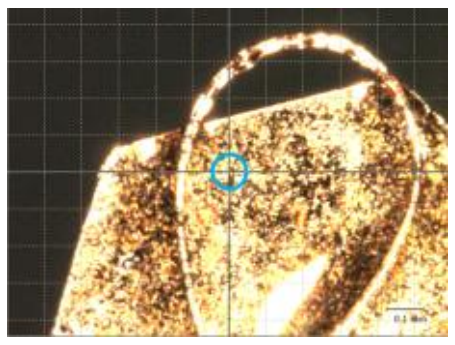 | 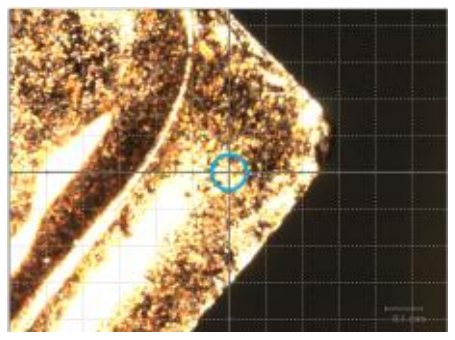 |
| Diff. down c* | 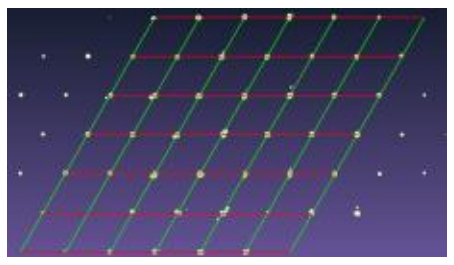 | 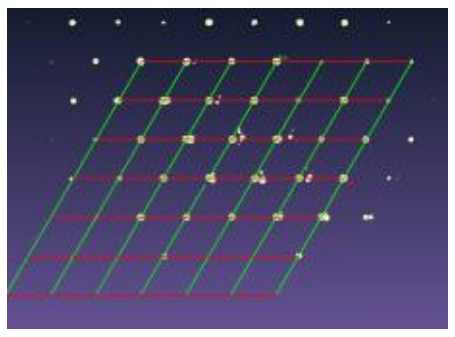 | 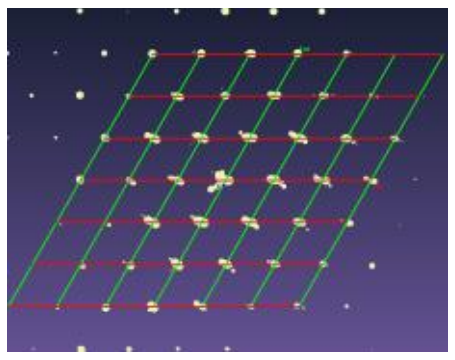 |
| Diff. down a* | 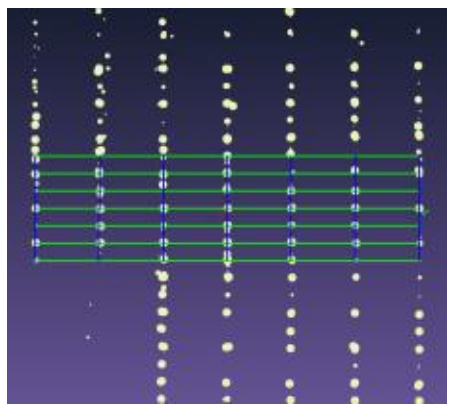 | 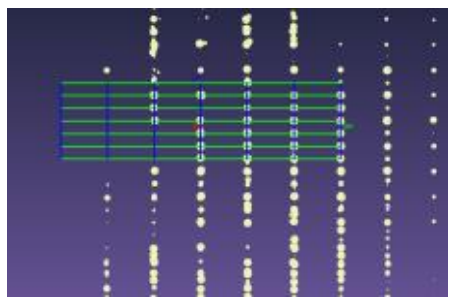 | 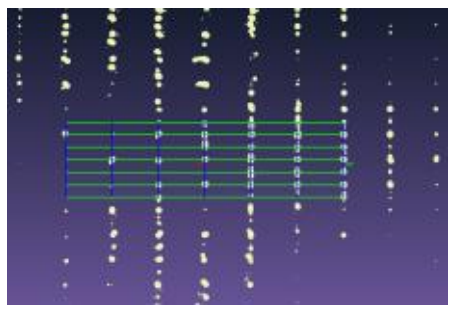 |
| c* histogram | 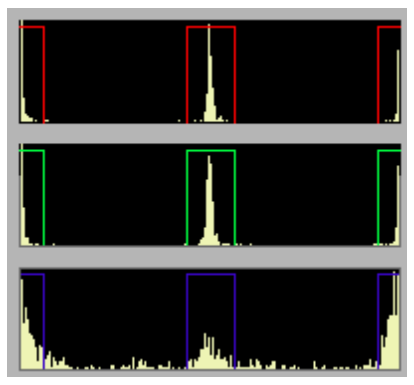 | 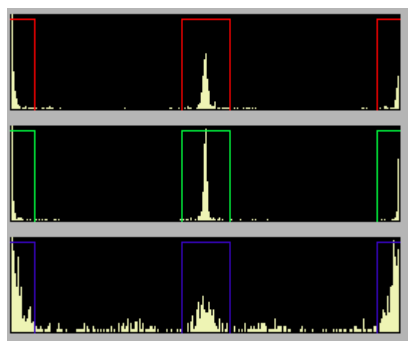 | 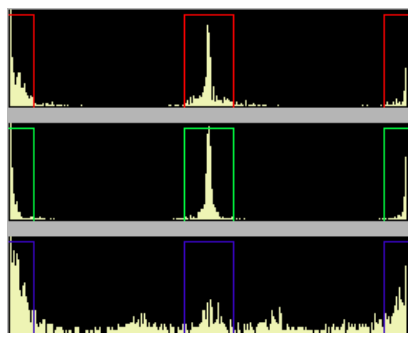 |
| Unit cell (Å, °) | a=3.4360, b=3.4487 c=12.682(3) V=130.15 Å³ | a=3.4420, b=3.4370 c=12.690(2) V=130.00 Å³ | a=3.444, b=3.440 c=12.696(3) V=130.54 Å³ |
| Indexed (%) | 524/706 (74%) | 606/856 (71%) | 612/1103 (56%) |

**Fig. S4. Single-crystal XRD screening data for xtl3 at three positions.** Row labels as in Fig. S2. The clean c*-axis diffraction and high fraction of indexed reflections (56–74%) confirm xtl3 as a high-quality 2H-$TaSe_2$ single crystal. This crystal was selected as the primary sample. Refined unit-cell parameters ($a \approx 3.44$ Å, $c \approx 12.69$ Å) are in excellent agreement with the literature 2H structure.

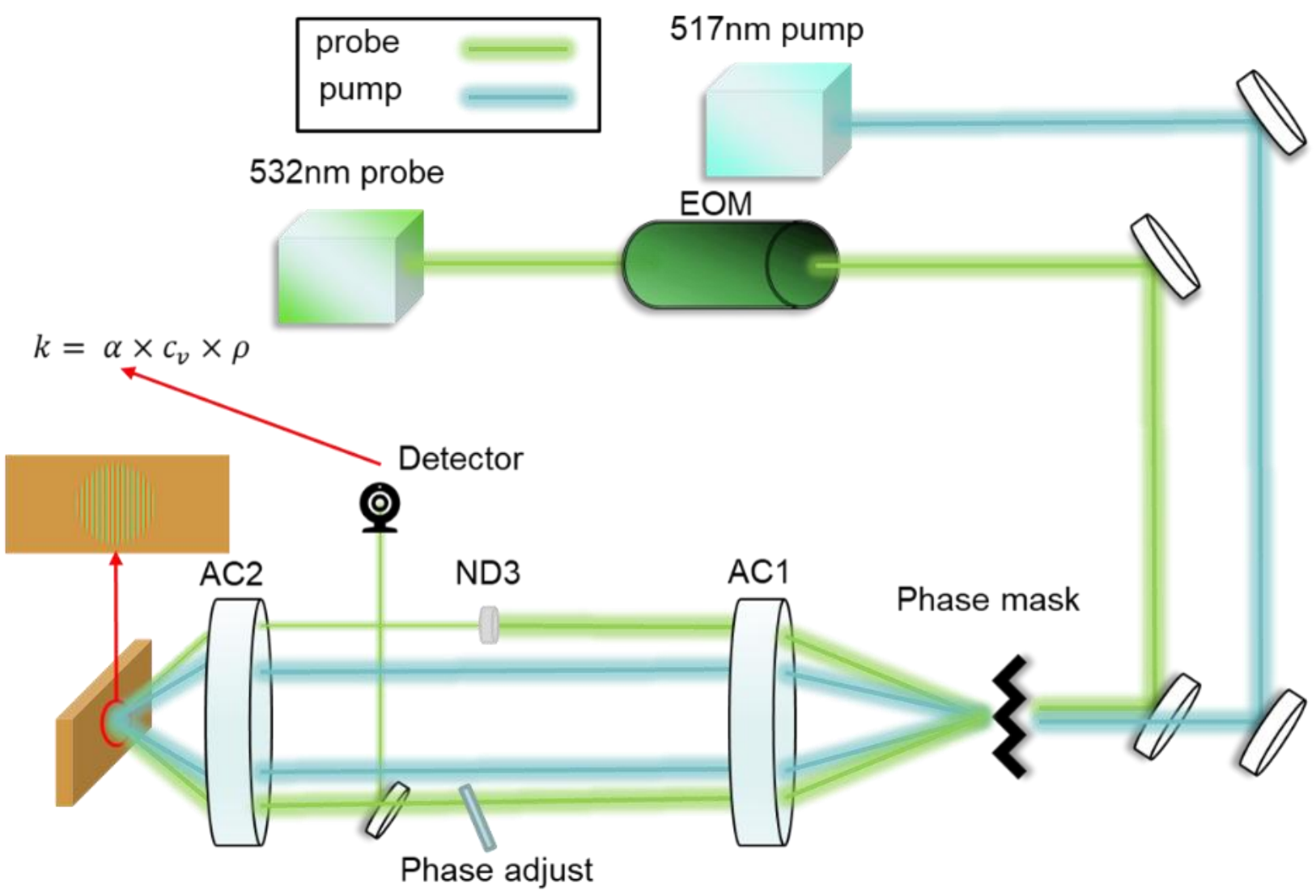

**Fig. S5. Schematic of the TTG optical setup in reflection geometry.** A phase mask splits the pump beam into two first-order beams that interfere at angle $\theta$ on the sample surface, creating a sinusoidal thermal grating of period $\Lambda$. The probe beam diffracts from the transient grating and is collected by a detector; a co-propagating reference beam enables heterodyne detection. An electro-optic modulator (EOM) controls the phase of the probe beam to enable heterodyne detection, allowing the in-phase (amplitude grating) and quadrature (phase grating) signal components to be recorded separately and improving the signal-to-noise ratio. A neutral density filter (ND) attenuates the pump beam to keep the peak surface temperature rise below ~10 K, ensuring the material response remains in the linear regime throughout the measurement.

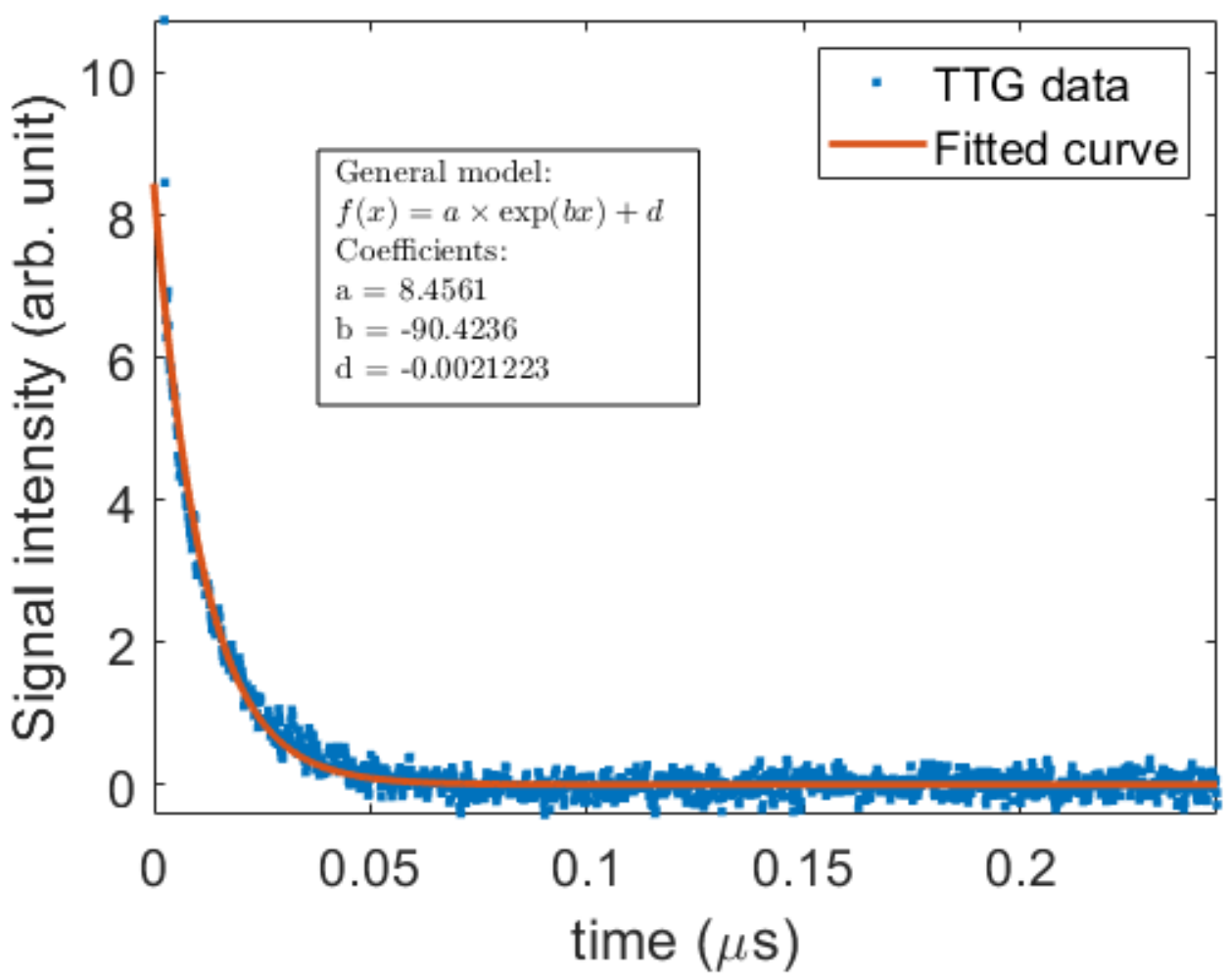

**Fig. S6. Fitting of thermal diffusivity from TTG data at room temperature.**

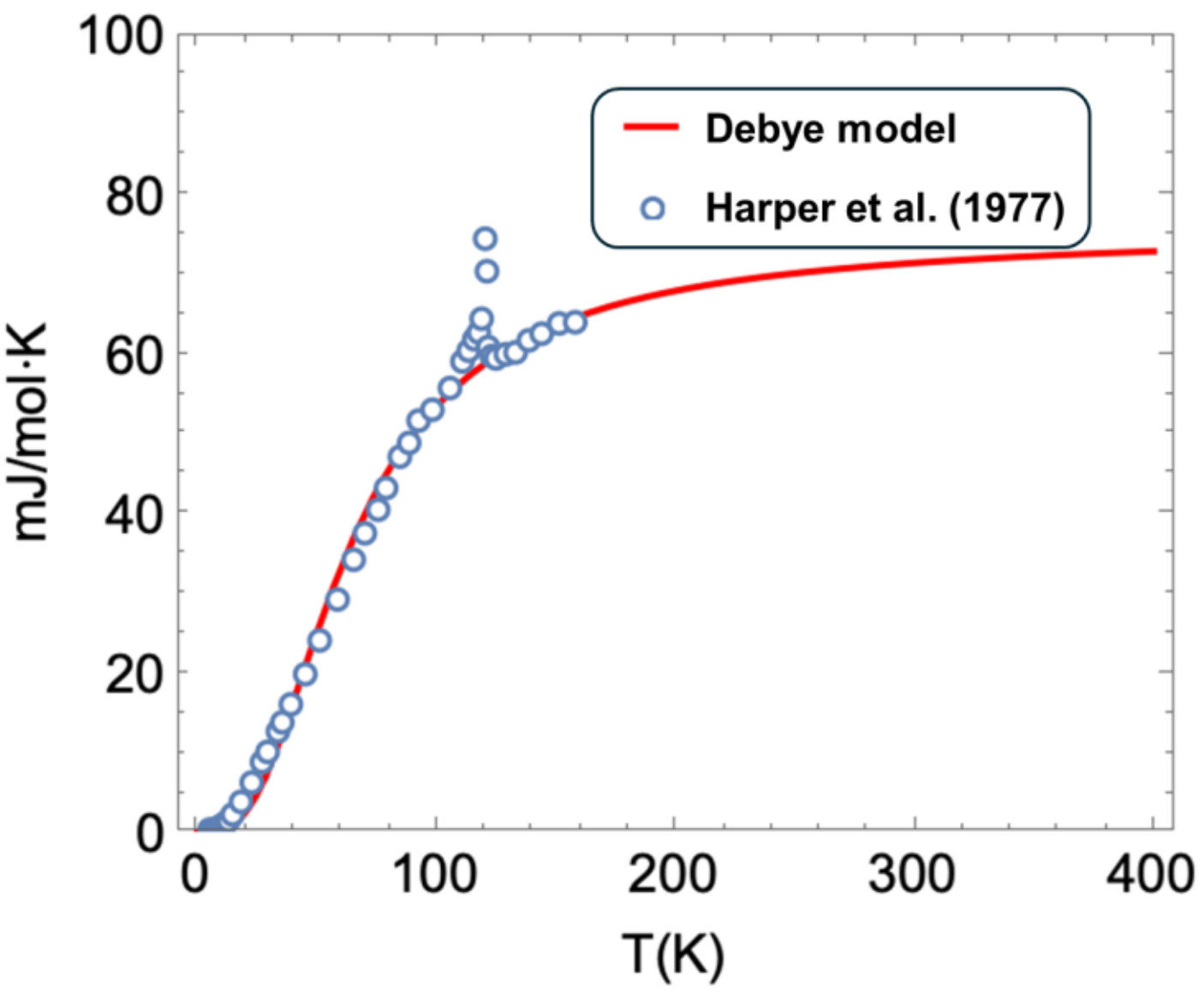

**Fig. S7. Temperature dependence of the specific heat capacity for 2H-TaSe₂.** The experimental data (blue circles), adapted from [47], exhibits a charge-density-wave (CDW) anomaly near 122 K. The solid red line represents the Debye model fit, which is utilized to extrapolate the specific heat values for temperatures beyond the experimental range ($T > 160\ K$).

**Table S1.**

| | 2H (n = 2) | 3R (n = 3, unstable) | 4H (n = 4) |
|---|---|---|---|
| Crystal system | Hexagonal | Trigonal/Rhombohedral | Hexagonal |
| Space group | $P6_3/mmc$ | R3m | $P6_3/mmc$ |
| a = b (Å) | 3.45 | 3.44 | 3.44 |
| c (Å) | 12.7 | 19.2 | 25.2 |
| V ($Å^3$) | ~130 | ~195 | ~260 |
| Z (#TaSe2 in unit cell) | 2 | 3 | 4 |

Known room-temperature polymorphs of $TaSe_2$ (all with trigonal prismatic $TaSe_6$ coordination). Z = number of formula units per unit cell. The 2H phase is the target polymorph for the present study.

**Table S2.**

| | xtl1 | xtl2 | xtl3 |
|---|---|---|---|
| Optical image | 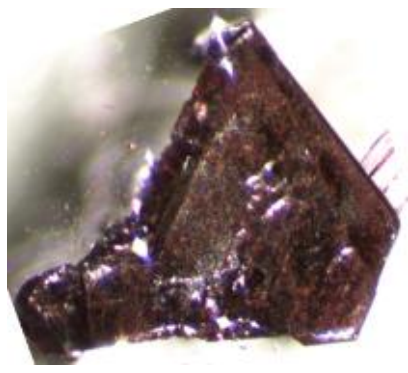 | 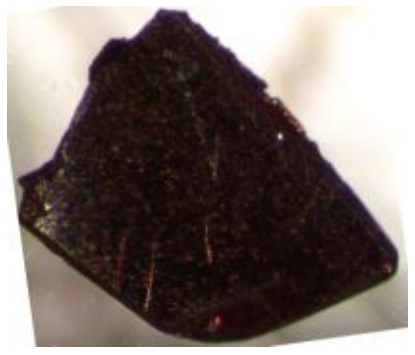 | 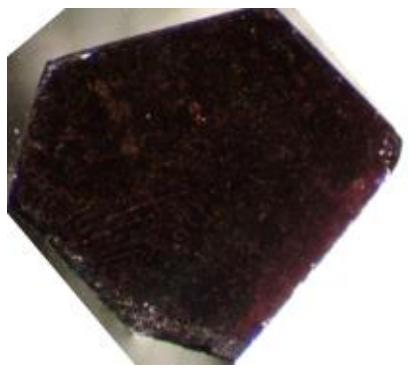 |
| Dimensions (mm) | 0.046 × 0.795 × 1.123 | 0.030 × 0.644 × 0.890 | 0.042 × 1.140 × 1.255 |
| Positions screened | 2 | 3 | 3 |
| % indexed (range) | 40–60% | 35–53% | 56–74% |
| c-axis quality | Poor–fair | Fair | Good |
| Use | Not mounted | Backup | Primary |

Summary of single-crystal XRD screening for the three $TaSe_2$ candidates. Percent indexed refers to the fraction of observed reflections that could be assigned to the 2H unit cell. Crystal xtl3 was selected as the primary sample.